\pdfoutput=1
\documentclass[reprint,showpacs,preprintnumbers,amsmath,amssymb,aps,nofootinbib]{revtex4-1}
\usepackage{ulem}
\usepackage{graphicx}% Include figure files
\usepackage{dcolumn}% Align table columns on decimal point
\usepackage[colorlinks=true, linkcolor = red, citecolor = blue]{hyperref}
\usepackage{subcaption}
\usepackage{natbib}
\usepackage{float}
\usepackage{comment}
\usepackage{braket}
\usepackage{xcolor}

\newcommand\JCH[1]{\textcolor{black}{#1}}

\begin{document}
%\title{Primordial Black Hole Formation for Scalar Field - Perfect Fluid Dominated Systems with Full General Relativity: The Case of a Scalar Field Dominated Universe}
\title{Primordial Black Hole Formation in a Scalar Field Dominated Universe}% Force line breaks with \\

\author{Ethan Milligan}
\email{e.milligan@qmul.ac.uk}
\affiliation{Astronomy Unit, Queen Mary University of London, Mile End Road, London, E1 4NS, UK.}

\author{Luis E. Padilla}
  \email{l.padilla@qmul.ac.uk}
\affiliation{Astronomy Unit, Queen Mary University of London, Mile End Road, London, E1 4NS, UK.}
  \author{David J. Mulryne}
  \email{d.mulryne@qmul.ac.uk}
  \affiliation{Astronomy Unit, Queen Mary University of London, Mile End Road, London, E1 4NS, UK.}
    \author{Juan Carlos Hidalgo}
  \email{hidalgo@icf.unam.mx}
  \affiliation{Instituto de Ciencias Físicas, Universidad Nacional Autónoma de México, 62210, Cuernavaca, Morelos, México.}
  
\date{\today}% 

\begin{abstract}
{We present a numerical code that solves the Misner-Sharp system for a spherically symmetric cosmological model containing both a scalar field and a perfect fluid. While the code is capable of exploring general scenarios involving a minimally coupled scalar field and perfect fluid, we focus on the regime where the scalar field dominates the dynamics, particularly in the post-inflationary scalar field-dominated scenario, where the universe is governed by a rapidly oscillating scalar field for a period lasting a few $e$-folds. We analyse the threshold for PBH formation under quadratic and quartic potentials, evolving configurations from superhorizon scales. Our results confirm that a quartic potential behavior is similar to the radiation-dominated universe, resulting in a PBH formation threshold close to the well-established value in radiation backgrounds. Conversely, in the quadratic case, we observe a significant deviation from the expected dust-like behaviour, due to wave-like effects opposing the gravitational collapse. While numerical limitations prevent us from evolving a wide range of initial conditions to determine a precise threshold for PBH formation, our findings suggest that PBH formation may be suppressed with respect to the pure dust scenario, allowing the formation of stable solitonic structures instead. This study highlights the importance of properly accounting for wave dynamics in oscillating scalar fields when characterising PBH formation.}\\
\end{abstract}

\maketitle

\section{Introduction}
The formation of primordial black holes (PBHs) in the early epochs of the universe was first conceived in 1968 by Zeldovich and Novikov \citep{ZeldovichNovikov} and then independently by Hawking \citep{Hawking71}. These seminal works suggested that large amplitude primordial overdensities can collapse under self-gravity to form black holes. As the interest towards PBH formation increased, initially due to PBHs possibly contributing to a significant part of dark matter \citep{Chapline:1975}, various formation mechanisms relevant for PBH formation have been explored \citep{Carr:2020}. These mechanisms include the formation of PBHs during inflation \citep{Clesse2015,Inomata2017,GARCIABELLIDO201747,Ezquiaga_2018}, the collision of bubbles from first order phase transitions \citep{Crawford:1982,HawkingMossStewart,KodamaHideoSasaki,Leach2000,Moss,Kitajima_2020,khlopov1998formationblackholesorder,Kawana_2022}, the collapse of cosmic strings \citep{TWBKibble_1976,HOGAN198487,HAWKING1989237,PolnarevAlexanderZembowicz,Garriga_1993,Caldwell_1996,blancopillado2021blackholescosmicstring,jenkins2020primordialblackholescusp,James_Turner_2020,Bertone_2020}, the collapse of domain walls produced during second order phase transitions \citep{dokuchaev2004quasarsformationclustersprimordial,rubin2000primordialblackholesnonequilibrium,Liu_2020}, the collapse of scalar condensate in the early universe \citep{Cotner_2017SUSY,Cotner_2017}, and specific baryogensis scenarios \citep{DolgovAlexandreSilk,Dolgov_2020,DOLGOV2009229,Kannike_2017}. 

Of all the mechanisms that have been proposed previously, the most widely studied is the collapse of overdense regions that are present in the early universe \citep{Nadezhin,Bicknell,Choptuik,EvansCharlesColeman,Niemeyer,Green,Polnarev_2012,Nakama_2014,Musco_2013,Yoo_2020} which may originate from inflationary quantum fluctuations. The PBH formation process was first explored in the context of a radiation-dominated phase by Carr \citep{Carr1975}. In this case, the threshold for collapse $\delta_{\rm th}$ is governed by non-zero pressure gradients and was originally calculated analytically to be $\delta_{\rm th}\approx1/3$. More recently, analytical and numerical studies have shown that the threshold depends on the initial shape of the overdensity and can range from $\delta_{\rm th} = 0.41 - 0.66$ \citep{NiemeyerJedamzik,Shibata_1999,IHawke_2002,Musco_2005,Polnarev_2007,HaradaTomohiro}.

Although the standard scenario of cosmic evolution assumes an instantaneous reheating, so that immediately after inflation the universe is radiation-dominated, this is only one of several possibilities and, in general, one may expect reheating to last for some time. In particular, the universe could enter a phase where a rapidly oscillating scalar field dominates the energy density of the universe, mimicking a dust-like phase, where copious PBH formation is expected. Scenarios that would invoke these phases of scalar field domination include a slow-reheating scenarios \citep{ABBOTT198229,Albrecht}, moduli fields \citep{Hagihara:2018uix, Banks:1993en}, or kination periods through quintessence models \citep{Kane:1993td, Pallis:2005bb}, among others.

In phases where a fast-oscillating scalar field dominates the early universe, the potential associated with the system is usually approximated by
\begin{equation}\label{eq:quadratic}
    V(\varphi) =  \frac{\lambda_n}{2n} \varphi^{2n},%\frac{\mu^2}{2}\varphi^2,
\end{equation}
where $\lambda_n$ is a free parameter of the model under consideration. In the case $n = 1$, $\lambda_1\equiv\mu^2$ can be interpreted as an effective mass parameter, whereas for the case $n = 2$, the parameter $\lambda_2 \equiv \lambda$ is interpreted as a self-interaction term. If the universe undergoes this scalar field dominated phase, its (time-averaged) equation of state can be determined by the power $n$ as \citep{PhysRevD.28.1243}:
\begin{equation}\label{eq:omega}
    \omega = \frac{n-1}{n+1}.
\end{equation}
This simplified description has typically led to approximating the condition for PBH formation using the formation criteria for universes dominated by perfect fluids. For example, in the case 
$n=1$, the formation criteria for matter-dominated scenarios -- such as conditions for the sphericity \citep{Harada:2016mhb}, angular momentum \citep{Harada:2017fjm}, or anisotropies \citep{Musco:2021sva} of the initial perturbations -- are typically adopted (see also \citep{1981SvA....25..406P,Khlopov:1980mg} for pioneer work), whereas for the case $n = 2$, the formation criterion for radiation domination is usualy adopted.

%where $\mu$ is an effective mass term of the inflaton field. During this time, the universe must have evolved (on average) similar to a universe filled with non-relativistic matter. 

%In the literature, different criteria have been considered to characterize the abundance of PBHs formed during this epoch. In the simplest scenario, results from a dust-dominated universe are often adopted, and from these, the relevant conclusions are drawn. For example, since for spherical perturbations the threshold for PBH formation in a dust-dominated universe is $\lim_{\omega \rightarrow 0}\delta_c \rightarrow \omega$ \citep{Harada:2013epa}, some authors have considered the only condition for PBH formation to be that reheating lasts long enough for the perturbations to grow and collapse (see for example \citep{Martin:2019nuw, Carrion:2021yeh}). However, even in this simple description, it has been shown that such a criterion would overestimate the expected abundance of PBHs, as effects like anisotropies \citep{Musco:2021sva}, the non-sphericity of the initial perturbation \citep{Harada:2016mhb}, or its angular momentum \citep{Harada:2017fjm} must play a significant role in preventing the gravitational collapse into 

The simplified assumption that the scalar field evolves like a perfect fluid does not account for the important fact that the wave-like dynamics and inherent oscillatory behaviour of the field can play a significant role. In particular, for the case $n=1$, it has been shown that this behaviour can prevent collapse and, instead, allow for the formation of primordial structures, such as solitonic- and halo-like configurations \citep{Niemeyer:2019gab,Eggemeier:2020zeg}. {This has been observed even in spherical symmetry \citep{PhysRevLett.72.2516}}. This wave-like property may also play an important role in discriminating the collapsing amplitudes, as discussed in \citep{Padilla:2021zgm,Hidalgo:2022yed, Padilla:2023lbv,Padilla:2024iyr,Padilla:2024cbq}. A question that remains open in this context is whether the perfect fluid approximation for an oscillating scalar field is valid to define collapsing amplitudes and thus determine the adequate formation time and mass of the resulting black hole. The goal of the present manuscript is to address the above question through a fully relativistic numerical treatment of the evolution of the oscillating scalar field.

Early hydrodynamical simulations of PBH formation \citep{Nadezhin,Bicknell,NovikovPolnerec} were based upon the Misner-Sharp formalism \citep{MisnerSharp}, which describes the gravitational collapse of a thermodynamic fluid under spherical symmetry in a curved spacetime. A number of works have used such a formalism to investigate critical behaviour \citep{NiemeyerJedamzik,Musco_2009,Musco_2013}, progressively improving the numerical method. Sasaki and Shibata \citep{Shibata_1999} highlighted the importance in constructing appropriate initial conditions excluding unphysical decaying modes. Musco \textit{et al} \citep{Musco_2005} investigated this further providing physical initial conditions for numerical simulations. Bloomfield \textit{et al} \citep{Bloomfield:2015} presented a comprehensive formalism for the numerical evolution of spherically symmetric perturbations employing the Misner-Sharp formalism, and building upon the ideas presented by the references mentioned above. 

Looking into the post-inflationary universe, the precise description of PBH formation demands a careful numerical treatment. With such motivation, we hereby introduce a code with the capability of exploring general cosmological scenarios involving the combination of a scalar field minimally coupled with a perfect fluid. Specifically, we extend the numerical formalism presented in \citep{Bloomfield:2015} for the perfect fluid case and produce a numerical code that solves the Misner-Sharp system for a universe containing both a scalar field and a perfect fluid. Following our motivation, we focus on the regime dominated by the scalar field. {To test the code, we} present two cases of interest: a post-inflationary universe dominated by a quadratic field and one dominated by a quartic field potential. For the first time, we simulate the full gravitational collapse of perturbations during this scalar field dominated scenario, leading to the formation of PBHs using the Misner-Sharp formalism.  

The paper is organized as follows. In Section~\ref{Sec:MS}, we begin by reviewing the Misner-Sharp formalism in detail. Section~\ref{Sec:msc} applies this formalism to cosmology, rewriting our system of equations in a more convenient form while also discussing the conditions necessary for PBH formation and the requirements for appropriate cosmological initial conditions. Section~\ref{Sec:IV} outlines the details of the numerical implementation used in our code, followed by Section~\ref{Sec:V}, where we present the numerical tests performed to validate our framework. The main results of this work are presented in Section~\ref{Sec:VI}, where we numerically evolve different initial conditions for scenarios in which either a quadratic or a quartic scalar field dominates the energy content of the universe. We find that the threshold for PBH formation in the quartic case is close to that of a radiation-dominated fluid. In the quadratic case, however, we were unable to fully evolve a wide range of initial conditions due to the challenges associated with tracking scalar field oscillations over a number of Hubble times. As a result, we could not determine a precise threshold for PBH formation and instead obtained only a constraint on this value. Finally, our conclusions are summarized in Section~\ref{Sec:VII}.

\section{Misner-Sharp Formalism for Scalar Field - Perfect fluid systems}\label{Sec:MS}

In this paper, we use the Misner-Sharp formalism \citep{MisnerSharp}, which uses the following line element of the spacetime
\begin{equation}\label{eq:MSmetric}
    ds^2 = -e^{2\phi}dt^2 + e^{\lambda}dA^2 + R^2d\Omega^2.
\end{equation}
The variables $\phi$, $\lambda$ and $R$ are functions of $A$ and $t$ only. The matter content is described by a scalar field plus a perfect fluid, with the stress-energy tensor given by
\begin{eqnarray}
    T_{\mu\nu} = &&(\rho_{\rm pf} + P_{\rm pf})u_{\mu}u_{\nu} + P_{\rm pf} g_{\mu\nu}  \nonumber \\
    &&- g_{\mu\nu}\left(\frac{1}{2}\partial_{\sigma}\varphi\partial^{\sigma}\varphi + V({\varphi})\right)+ \partial_{\mu}\varphi\partial_{\nu}\varphi.
\end{eqnarray}
Here, $V(\varphi)$ is the potential of the scalar field $\varphi$, while $\rho_{\rm pf}$, $P_{\rm pf}$, and $u_\mu$ represent the energy density, pressure, and four velocity of the additional perfect fluid, respectively.

In the following, we will work in a coordinate system where the coordinates are comoving with the perfect fluid. In this system, the four-velocity of the perfect fluid is given by
\begin{equation}\label{eq:comovil}
    u^t = e^{-\phi}, \ \ \ u^i = 0; \ i = r,\theta, \phi.
\end{equation}
In other words, we fix the foliation by assuming that the normal to the hypersurfaces coincides with the fluid 4-velocity. The nontrivial time-time, time-space and space-space components of the Einstein equation are calculated as follows:
\begin{subequations}
\begin{eqnarray}\label{eq:rhot}
        8\pi e^{2\phi}\rho_{\rm T} =&& \frac{1}{R^2}\bigg(e^{2\phi} + \Dot{R}^2 - e^{2\phi - \lambda} (R^{\prime})^2\bigg)\nonumber \\
        &&+ \frac{e^{-\lambda}}{R}\bigg(e^{\lambda}\Dot{\lambda}\Dot{R} + e^{2\phi}R^{\prime}\lambda^{\prime} - 2e^{2\phi}R^{\prime\prime}\bigg) ,
\end{eqnarray}
\begin{eqnarray}
    8\pi e^{\lambda} P_{\rm T} = &&\frac{1}{R^2}\bigg[(R^{\prime})^2 + 2RR^{\prime}\phi^{\prime} \nonumber - e^{\lambda - 2\phi}\\ 
    &&
    \times \bigg(e^{2\phi} + \Dot{R}- 2R\Dot{R}\Dot{\phi} + 2R\Ddot{R}\bigg)\bigg] ,\label{eq:pt}
\end{eqnarray}
\begin{equation}\label{eq:cross}
    8\pi\Dot{\varphi}\varphi^{\prime} = \frac{\Dot{\lambda}R^{\prime} + 2\Dot{R}\phi^{\prime} - 2\Dot{R}^{\prime}}{R} .
\end{equation}
\end{subequations}
In the above equations, dots and primes denote derivatives with respect to $t$ and $A$, respectively. We have also defined the total energy density $\rho_{\rm T}$ and total pressure $P_{\rm T}$ of the system, as $\rho_{\rm T} = \rho_{\rm pf} + \rho_\varphi$ and $P_{\rm T} = P_{\rm pf} + P_\varphi$, where the energy density and pressure associated with the scalar field are given by
\begin{subequations}
\begin{equation}\label{eq:sfed}
    \rho_\varphi = \frac{e^{-2\phi}}{2}\Dot{\varphi}^2 + \frac{e^{-\lambda}}{2}(\varphi^{\prime})^2 + V(\varphi),
\end{equation}
\begin{equation}
    P_\varphi = \frac{e^{-2\phi}}{2}\Dot{\varphi}^2 + \frac{e^{-\lambda}}{2}(\varphi^{\prime})^2 - V(\varphi).
\end{equation}
\end{subequations}
The term on the left-hand side of \eqref{eq:cross} is a momentum density and can be understood as an energy flux associated with the scalar field. %Therefore, this term serves as a key distinction from the simpler case of considering only perfect fluids.

%This equation will often be rewritten to remove $\Dot{\lambda}$ dependencies in the evolution equations. The term on the right-hand side of the equation is an obvious difference from the purely radiation case. 

Using the conservation of the stress-energy tensor, $\nabla_\mu T^{\mu\nu} = 0$, we derive the energy and momentum conservation equations for the perfect fluid, as well as the Klein-Gordon (KG) equation of motion for the scalar field:  
\begin{subequations}
\begin{equation}\label{eq:drho}
    2R\Dot{\rho}_{\rm pf} + (P_{\rm pf} + \rho_{\rm pf})(4\dot{R} + \dot{\lambda}R) = 0,
\end{equation}
\begin{equation}\label{eq:phi}
    P_{\rm pf}^{\prime} + (P_{\rm pf} + \rho_{\rm pf})\phi^{\prime} = 0,
\end{equation}
\begin{eqnarray}
    -e^{-2\phi}\bigg(\frac{2\Dot{R}\Dot{\varphi}}{R} + \frac{1}{2}\Dot{\lambda}\Dot{\varphi} - 2\Dot{\phi}\Dot{\varphi} + \Ddot{\varphi}\bigg) &&\nonumber\\
    + e^{-\lambda}\bigg(\frac{2R^{\prime}\varphi^{\prime}}{R} - \frac{1}{2}\lambda^{\prime}\varphi^{\prime} + \phi^{\prime}\varphi^{\prime} + \varphi^{\prime\prime}\bigg) &&= V(\varphi)_{,\varphi},
\end{eqnarray}
\end{subequations}
where $V(\varphi)_{,\varphi}$ is the derivative of the scalar field potential with respect to $\varphi$. We emphasise that no interaction between the scalar field and the perfect fluid is considered here.

To recast the KG equation as a first order system, we define the following auxiliary variables
\begin{equation}
    \chi \equiv \varphi^{\prime} , \ \ \ \Pi \equiv \Dot{\varphi}e^{-\phi}.
\end{equation}
With which we recast the KG equation as the system:
\begin{subequations}\label{eq:KG}
\begin{equation}
    \dot\varphi = \Pi e^\phi,
\end{equation}
\begin{equation}
    \dot\chi = (\Pi^{'}+\phi^{'}\Pi)e^\phi,
\end{equation}
\begin{eqnarray}
    -e^{-\phi}\bigg(\frac{2\Dot{R}}{R}\Pi + \frac{\Dot{\lambda}}{2}\Pi  + \Dot{\Pi}\bigg)&&\nonumber \\ 
    + e^{-\lambda}\bigg[\bigg(\frac{2R^{\prime}}{R} + \phi^{\prime} -\frac{\lambda^{\prime}}{2}\bigg)\chi + \chi^{\prime}\bigg] &&= V(\varphi)_{,\varphi}.
\end{eqnarray}
\end{subequations}
Hereafter, we will express the equations in terms of these auxiliary variables for the scalar field.

We define the total local mass of the system $m_{\rm T}$ as:
\begin{equation}\label{eq:mT}
    m_{\rm T} = m_{\rm pf} + m_\varphi = 4\pi\int^{A}_0(\rho_{\rm pf}+\rho_\varphi)R^2R^{\prime}dA.
\end{equation}
The quantity $m_{\rm T}$ {not necessary coincides with} the Misner-Sharp mass, which is a measure of the total energy (mass) contained within the radius $A$. We can calculate the Misner-Sharp mass from the definition
\begin{equation}\label{eq:MSmass}
    M_{\rm MS} = \frac{R}{2}(1 - \nabla^aR\nabla_aR),
\end{equation}
with $\nabla^aR\nabla_aR = \Gamma^2 - U^2$, where we have used the following variables:
\begin{equation}\label{eq:ugam}
    U = e^{-\phi}\Dot{R},\ \ \  \Gamma = e^{-\lambda/2}R^{\prime}.
\end{equation}
Here $U$ represents a coordinate velocity of the system at comoving radius $A$, and $\Gamma$ is an alternative to $\lambda$. The following related expressions will be useful in simplifying future equations:
\begin{subequations}
\begin{equation}\label{eq:rddot}
    e^{-\phi}\Ddot{R} = \Dot{U} + U\Dot{\phi},
\end{equation}
\begin{equation}\label{eq:Rbp}
    e^{-\lambda/2}R^{\prime\prime} = \Gamma^{\prime} + \frac{1}{2}\Gamma\lambda^{\prime},
\end{equation}
\begin{equation}\label{eq:mpf}
    m_{\rm pf}^{\prime} = 4\pi R^2R^{\prime}\rho_{\rm pf},
\end{equation}
\begin{equation}\label{eq:mvphi}
    m_\varphi^{\prime} = 4\pi R^2R^{\prime}\rho_\varphi.
\end{equation}
\end{subequations}

Before moving to the cosmological description of the system, let us rewrite the above equations in a more convenient form. First, we derive suitable constraint equations. The constraint for $\phi$ is obtained by rearranging Eq.~\eqref{eq:phi} as:
\begin{equation}\label{eq:phiprime}
     -\frac{P_{\rm pf}^{\prime}}{P_{\rm pf} + \rho_{\rm pf}} = \phi^{\prime}.
\end{equation}
To derive the second constraint equation for $\Gamma$, we start with Eq.~\eqref{eq:rhot} and substitute in Eqs.~\eqref{eq:ugam}, \eqref{eq:Rbp} and \eqref{eq:cross}, yielding:
\begin{equation}\label{eq:gam2}
    \Gamma^2 = 1 + U^2 - \frac{2m_{\rm T}}{R} + \frac{2 I_1}{R},
\end{equation}
where 
\begin{equation}\nonumber
    I_1 = 4\pi \int^A_0 R^2 U \chi\Pi dA.
\end{equation}
Note that the scalar field component introduces an integral term into the constraint equation, which is related to the energy flux of the field. Then, the Misner-Sharp mass defined in Eq.~\eqref{eq:MSmass} is expressed simply as $M_{\rm MS} = m_{\rm T} - I_1$. We complement the set of constraint equations with those given in Eqs.~\eqref{eq:mpf} and \eqref{eq:mvphi}.

Next, we derive the three necessary evolution equations for $R$, $m_{\rm pf}$, and $U$, which, together with the KG equation, Eq.~\eqref{eq:KG}, constitute the complete system of differential equations for our scalar field - perfect fluid system. We first derive the evolution equation for the perfect fluid mass by multiplying Eq.~\eqref{eq:drho} by $2\pi RR^{\prime}$ and substituting in Eq.~\eqref{eq:cross}. After rearranging and simplifying, we obtain:
\begin{eqnarray}
    &&\frac{d}{dt}(4\pi\rho_{\rm pf} R^2R^{\prime}) + 4\pi\frac{d}{dA}(P_{\rm pf} R^2\Dot{R}) \nonumber \\
    = && 
    - 16\pi^2 (\rho_{\rm pf} + P_{\rm pf})e^{\phi}R^3\chi\Pi.
\end{eqnarray}
Integrating over the radial coordinate from $0$ to $A$, and using Eq.~\eqref{eq:ugam} and the perfect fluid part of Eq.~\eqref{eq:mT}, the final expression becomes:
\begin{equation}
    \Dot{m}_{\rm pf} + 4\pi e^{\phi} R^2 P_{\rm pf} U = 4\pi I_2,
\end{equation}
where 
\begin{equation}\nonumber
    I_2 = 4\pi \int^A_0 (\rho_{\rm pf} + P_{\rm pf})e^{\phi}R^3\chi\Pi dA.
\end{equation}

To determine the evolution equation for $U$, we substitute Eqs.~\eqref{eq:ugam} and \eqref{eq:rddot} into Eq.~\eqref{eq:pt}, which yields:
\begin{equation}
    2e^{-\phi}R\Dot{U} = \Gamma^2 - 1- U^2 - 8\pi P_{\rm T} R^2 + \frac{2\Gamma^2R\phi^{\prime}}{R^{\prime}},
\end{equation}
and can be further simplified using Eqs.~\eqref{eq:phiprime} and \eqref{eq:gam2}:
\begin{equation}
    \Dot{U} =-e^{\phi}\bigg(\frac{\Gamma^2 P_{\rm pf}^{\prime}}{R^{\prime}(\rho_{\rm pf} + P_{\rm pf})}+\frac{4\pi P_{\rm T} R^3+m_{\rm T}  - I_1}{R^2}
    \bigg).
\end{equation}
The evolution equation for $R$ is already given by the left-hand side of Eq.~\eqref{eq:ugam}.

Let us summarise the Misner-Sharp equations for our complete scalar field - perfect fluid system for convenience (evolution plus constraint equations):
\begin{subequations}\label{eq:MS}
\begin{equation}
    \dot\varphi = \Pi e^\phi,
\end{equation}
\begin{equation}
    \dot\chi = (\Pi^{'}+\phi^{'}\Pi )e^\phi,
\end{equation}
\begin{eqnarray}
    -e^{-\phi}\bigg(\frac{2\Dot{R}}{R}\Pi + \frac{\Dot{\lambda}}{2}\Pi  + \Dot{\Pi}\bigg)&&\nonumber \\ 
    + e^{-\lambda}\bigg[\bigg(\frac{2R^{\prime}}{R} + \phi^{\prime} -\frac{\lambda^{\prime}}{2}\bigg)\chi + \chi^{\prime}\bigg] &&= V(\varphi)_{,\varphi},
\end{eqnarray}
\begin{equation}\label{eq:Rdot}
    \Dot{R} = U e^{\phi},
\end{equation}
\begin{equation}\label{eq:mpfdot}
    \Dot{m}_{\rm pf} + 4\pi e^{\phi} R^2 P_{\rm pf} U = 4\pi I_2,
\end{equation}
\vspace{-0.25cm}
\begin{equation}\label{Udot}
\Dot{U} =-e^{\phi}\bigg(\frac{\Gamma^2 P_{\rm pf}^{\prime}}{R^{\prime}(\rho_{\rm pf} + P_{\rm pf})}+\frac{4\pi P_{\rm T} R^3+m_{\rm T}  - I_1}{R^2}
    \bigg),
\end{equation}
\begin{equation}\label{eq:rhopf}
    -\frac{P_{\rm pf}^{\prime}}{P_{\rm pf} + \rho_{\rm pf}} = \phi^{\prime},
\end{equation}
\begin{equation}\label{eq:rhosf}
    \rho_{\rm pf} = \frac{m^{\prime}_{\rm pf}}{4\pi R^2R^{\prime}},
\end{equation}
\begin{equation}
    \rho_\varphi = \frac{m^{\prime}_\varphi}{4\pi R^2R^{\prime}},
\end{equation}
\begin{equation}
    \Gamma^2 = 1 + U^2 - \frac{2m_{\rm T}}{R} + \frac{2I_1}{R}.
\end{equation}
\end{subequations}
%For later convenience, we also include the following expression\lp{I don't remember. Why this expression is important?  - I use it for the derivation of equations in the Misner Hernandez formalism. I do not remember using it in the Misner Sharp formalism. I think we can remove this sentence or entire equation.}
%\begin{equation}
%    \dot{\rho}_{\rm pf} = -(\rho_{\rm pf}+P_{\rm pf})e^{\phi}\bigg(2\frac{U}{R} + \frac{U^{\prime}}{R^{\prime}} + \frac{4\pi R\chi\Pi}{R^{\prime}}\bigg).
%\end{equation}

To complete the Misner-Sharp formalism, we must specify boundary conditions. To make the physical quantities consistent, we add the following boundary conditions: $R(t,0)=0$, $U(t,0)=0$, and $m(t,0)=0$. By spherical symmetry, $\rho^{\prime}(t,0)=P^{\prime}(t,0)=\chi(t,0)=\Pi^{\prime}(t,0)=0$ and thus $\phi^{\prime}(t,0)=0$. From L'Hôpital's rule we find that $\Gamma^2(t,A\rightarrow0)=1$.

\section{PBH formation through the Misner-Sharp formalism}\label{Sec:msc}

\subsection{Cosmological Variables}

As expected, the Misner-Sharp formalism can reproduce the FLRW background universe (see Appendix \ref{app:background}). Taking advantage of this, it is useful to factor out the background solution from the variables in Eqs.~\eqref{eq:MS} to enhance accuracy and numerical stability. We thus take,
\begin{subequations}\label{eqs:cosmoMD}
\begin{equation}
    R = aA\tilde{R},
\end{equation}
\begin{equation}
    \rho_{\rm pf} = \rho_{\rm T,b}\tilde{\rho}_{\rm pf},~~~~~\rho_\varphi = \rho_{\rm T,b}\tilde{\rho}_\varphi,~~~~~\rho_{\rm T} = \rho_{\rm T,b}\tilde{\rho}_{\rm T},
\end{equation}
\begin{equation}
    P_{\rm pf}=\rho_{\rm T,b}\tilde{P}_{\rm pf}, ~~~~~ P_\varphi=\rho_{\rm T,b}\tilde{P}_\varphi,~~~~~P_{\rm T}=\rho_{\rm T,b}\tilde{P}_{\rm T}, 
\end{equation}
\begin{equation}
    m_{\rm pf} = \frac{4\pi}{3}\rho_{\rm T,b}R^3\tilde{m}_{\rm pf},~~~~~  m_{\rm T} = \frac{4\pi}{3}\rho_{\rm T,b}R^3\tilde{m}_{\rm T},
\end{equation}
\begin{equation}
    U = HR\tilde{U},
\end{equation}
\begin{equation}
    \varphi = R_H\sqrt{\rho_{\rm T,b}}\tilde{\varphi},
\end{equation}
\begin{equation}
    \chi = \sqrt{\rho_{\rm T,b}}\tilde{\chi},
\end{equation}
\begin{equation}
    \Pi = \sqrt{\rho_{\rm T,b}}\tilde{\Pi}.
\end{equation}
\end{subequations}
In the above equations, $a$, $H$, and $\rho_{\rm T,b}$ represent the scale factor, Hubble parameter, and mean total background energy density, respectively. The dimensionless variables represent a rescaling of physical variables with respect to their expected values in a homogeneous FRLW universe. This approach allows us to isolate and track deviations from the background cosmology more effectively. Furthermore, we assume that at the background level, an effective equation of state $w$ can be defined such that the total mean background pressure is written as $P_{\rm T,b} = w\rho_{\rm T,b}$. We also introduce $R_H$, the Hubble horizon radius at the beginning of the evolution. Using these new tilde variables, Eqs.~\eqref{eq:MS} are rewritten as follows:
\begin{widetext}
\begin{subequations}
\begin{equation}
    \frac{1}{H}\dot{\tilde{\varphi}} = \frac{e^{\phi}}{HR_H}\tilde{\Pi} + \frac{\tilde{\varphi}}{\alpha},
\end{equation}
\begin{equation}
    \frac{1}{H}\dot{\tilde{\chi}} = \frac{e^{\phi}}{H}(\tilde{\Pi}^{\prime} + \phi^{\prime}\tilde{\Pi}) + \frac{\tilde{\chi}}{\alpha},
\end{equation}
\begin{equation}
    \frac{1}{H}\Dot{\tilde{\Pi}} = \frac{\tilde{\Pi}}{\alpha} +e^{\phi}\bigg[\frac{\Gamma^2}{a^2H^2(\tilde{R} + A\tilde{R}^{\prime})^2}\bigg[\bigg(\frac{2(A\tilde{R})^{\prime}}{A\tilde{R}} + \phi^{\prime} -\frac{1}{2}\lambda^{\prime}\bigg)\tilde{\chi} + \tilde{\chi}^{\prime}\bigg] - \bigg(2\tilde{U} + \frac{(A\tilde{R}\tilde{U})^{\prime} + \frac{3}{2}HA\tilde{R}\tilde{\chi}\tilde{\Pi}}{(A\tilde{R})^{\prime}}\bigg)\tilde{\Pi} - \frac{V(\tilde{\varphi})_{,\tilde{\varphi}}}{\rho_{\rm T,b} H R_H}\bigg],
\end{equation}
\begin{equation}
   \frac{1}{H}\dot{\tilde{R}} =  \tilde{R}(e^{\phi}\tilde{U} - 1),
\end{equation}
\begin{equation}
    \frac{1}{H}\Dot{\tilde{m}}_{\rm pf} = \frac{2}{\alpha}\tilde{m}_{\rm pf} - 3\tilde{U}e^{\phi}(\tilde{P}_{\rm pf} + \tilde{m}_{\rm pf}) + \frac{9H}{2A^3\tilde{R}^3}\tilde I_2,
\end{equation}
\begin{equation}
     \frac{1}{H}\Dot{\tilde{U}}= \frac{\tilde{U}}{\alpha} - e^{\phi}\bigg[\Gamma^2\frac{P^{\prime}_{\rm pf}}{H^2(aA\tilde{R})(aA\tilde{R})^{\prime}(\tilde{\rho}_{\rm pf} + \tilde{P}_{\rm pf})} + \frac{1}{2}(2\tilde{U}^2 + \tilde{m}_{\rm T} + 3\tilde{P}_{\rm T}) - \frac{3H}{2A^3\tilde{R}^3}\tilde I_{1}\bigg],
\end{equation}
\begin{equation}
   \phi^{\prime} =  -\frac{\tilde{P}^{\prime}_{\rm pf}}{\tilde{P}_{\rm pf} + \tilde{\rho_{\rm pf} }},
\end{equation}
\begin{equation}
     \tilde{\rho}_{\rm pf} = \tilde{m}_{\rm pf}  + \frac{A\tilde{R}}{3(A\tilde{R})^{\prime}}\tilde{m}^{\prime}_{\rm pf},
\end{equation}
\begin{equation}
     \tilde{\rho}_{\varphi} = \tilde{m}_{\varphi}  + \frac{A\tilde{R}}{3(A\tilde{R})^{\prime}}\tilde{m}^{\prime}_{\varphi},
\end{equation}
\begin{equation}\label{eq:Gsquared}
    \begin{split}
    \Gamma^2 = 1 + H^2(aA)^2\tilde{R}^2(\tilde{U}^2 - \tilde{m}_{\rm T}) + \frac{3a^2H^3}{A\tilde{R}}\tilde I_1,
    \end{split}
\end{equation}
\end{subequations}
\end{widetext}
where 
\begin{equation}
    \tilde I_{1} = \int^A_0\tilde{U}A^3\tilde{R}^3\tilde{\chi}\tilde{\Pi} dA,\nonumber 
\end{equation}
\begin{equation}
    \tilde I_2 = \int^A_0(\tilde{\rho}_{\rm pf}+\tilde{P}_{\rm pf})A^3\tilde{R}^3e^{\phi}\tilde{\chi}\tilde{\Pi}dA.\nonumber 
\end{equation}
In the above equations, we have defined $\alpha = 2/3(1+w)$. 

For numerical evolution, the process slows significantly as time progresses due to the $1/t$ term in the equations of motion. To avoid this, it is useful to work in terms of a dimensionless time variable. In particular, we choose to evolve in logarithmic time:
\begin{equation}
    \xi = \ln\bigg(\frac{t}{t_0}\bigg) = \frac{1}{\alpha} \ln a ,
\end{equation}
allowing for quantities that depend on powers of $a$ to evolve linearly. Consequently, we define the following useful relations
\begin{equation}
    \partial_t = \frac{H}{\alpha}\partial_\xi.
\end{equation}
The scale factor and Hubble parameter are expressed as 
\begin{equation}
    a = e^{\alpha \xi}
\end{equation}
\begin{equation}
    H = \frac{\alpha}{t_0}e^{-\xi} = \frac{1}{R_H}e^{-\xi}.
\end{equation}
Furthermore, it is convenient to define $A\equiv\Bar{A}R_H$, where $\Bar{A}$ is dimensionless. Rewriting the cosmological Misner-Sharp equations in terms of logarithmic time and this dimensionless spatial coordinate, we obtain the following evolution equations:
\begin{widetext}
\begin{subequations}\label{eq:final}
\begin{equation}\label{eq:SFevo}
    \partial_\xi\tilde{\varphi} = \alpha e^{\phi+\xi}\tilde{\Pi} + \tilde{\varphi},
\end{equation}
\begin{equation}\label{eq:chifinal}
    \partial_\xi\tilde{\chi} =  \alpha e^{\phi + \xi}(\tilde{\Pi}^{\prime} + \phi^{\prime}\tilde{\Pi}) + \tilde{\chi},
\end{equation}
\begin{equation}\label{eq:SFPIevo}
    \partial_\xi\tilde{\Pi}=\tilde{\Pi} +\alpha e^{\phi}\bigg[\frac{\Bar{\Gamma}^2 e^{-\xi}}{(\tilde{R} + \Bar{A}\tilde{R}^{\prime})^2}\bigg[\bigg(\frac{2(\Bar{A}\tilde{R})^{\prime}}{\Bar{A}\tilde{R}} + \phi^{\prime} -\frac{1}{2}\lambda^{\prime}\bigg)\tilde{\chi} + \tilde{\chi}^{\prime}\bigg] - \bigg(2\tilde{U} + \frac{(A\tilde{R}\tilde{U})^{\prime} + \frac{3}{2}e^{-\xi}\Bar{A}\tilde{R}\tilde{\chi}\tilde{\Pi}}{(A\tilde{R})^{\prime}}\bigg)\tilde{\Pi}- e^{\xi}\tilde{V}(\tilde{\varphi})_{,\tilde{\varphi}}\bigg],
\end{equation}
\begin{equation}\label{eq:Rfinal}
    \partial_\xi\tilde{R} = \alpha\tilde{R}(\tilde{U}e^{\phi} - 1),
\end{equation}
\begin{equation}\label{eq:mfinal}
    \partial_\xi\tilde{m}_{\rm pf} = 3\alpha\bigg[(1+w)\tilde{m}_{\rm pf} - \tilde{U}e^{\phi}(\tilde{P}_{\rm pf} + \tilde{m}_{\rm pf}) + \frac{3e^{-\xi}}{2\Bar{A}^3\tilde{R}^3}\mathcal{I}_2\bigg],% \int^{\Bar{A}}_0(\tilde{\rho}_r+\tilde{P}_r)\Bar{A}^3\tilde{R}^3e^{\phi}\tilde{\chi}\tilde{\Pi}d\Bar{A}
\end{equation}
\begin{equation}\label{eq:Ufinal}
     \partial_\xi\tilde{U}= \tilde{U} - \alpha e^{\phi}\bigg[\Bar{\Gamma}^2\frac{P^{\prime}_{\rm pf}}{(\Bar{A}\tilde{R})(\Bar{A}\tilde{R})^{\prime}(\tilde{\rho}_{\rm pf} + \tilde{P}_{\rm pf})} + \frac{1}{2}(2\tilde{U}^2 + \tilde{m}_{\rm T} + 3\tilde{P}_{\rm T}) - \frac{3e^{-\xi}}{2\Bar{A}^3\tilde{R}^3}\mathcal{I}_1\bigg], %\int^{\Bar{A}}_0\tilde{U}\bar{A}^3\tilde{R}^3\tilde{\chi}\tilde{\Pi} d\Bar{A}
\end{equation}
\begin{equation}\label{eq:lapseint}
    \phi^{\prime} = -\frac{\tilde{P}^{\prime}_{\rm pf}}{\tilde{\rho}_{\rm pf} + \tilde{P}_{\rm pf}},
\end{equation}
\begin{equation}\label{eq:rhoFinal}
    \tilde{\rho}_{\rm pf} = \tilde{m}_{\rm pf}  + \frac{\Bar{A}\tilde{R}}{3(\Bar{A}\tilde{R})^{\prime}}\tilde{m}^{\prime}_{\rm pf},
\end{equation}
\begin{equation}
    \tilde{\rho}_{\varphi} = \tilde{m}_{\varphi}  + \frac{\Bar{A}\tilde{R}}{3(\Bar{A}\tilde{R})^{\prime}}\tilde{m}^{\prime}_{\varphi},
\end{equation}
\begin{equation}\label{eq:Gamma}
    \Bar{\Gamma}^2 = \frac{\Gamma^2}{a^2H^2R_H^2} = e^{2(1-\alpha)\xi} +\Bar{A}^2\tilde{R}^2(\tilde{U}^2 - \tilde{m}_{\rm T}) + \frac{3e^{-\xi}}{\Bar{A}\tilde{R}}\mathcal{I}_1, %\int^{\Bar{A}}_0\tilde{U}\Bar{A}^3\tilde{R}^3\tilde{\chi}\tilde{\Pi}d\Bar{A} 
\end{equation}

\end{subequations}
\end{widetext}
where we normalize the integrals as
\begin{equation}\nonumber
    \mathcal{I}_1 = \frac{\tilde{I}_1}{R_H^ 4}, \ \ \ \ \ \ \mathcal{I}_2 = \frac{\tilde{I}_2}{R_H^ 4}.
\end{equation}
Note that, in the above system, we repurposed primes to denote derivatives with respect to $\Bar{A}$. 

The set of Eqs.~\eqref{eq:final} constitutes the complete dimensionless cosmological Misner-Sharp equations of motion, which have been implemented into the simulation code. These equations are supplemented by the inner boundary conditions $\tilde{\rho}_{\rm pf}^{\prime} = \tilde{R}^{\prime} = \tilde{m}_{\rm pf}^{\prime} = \tilde{U}^{\prime} = \tilde{\chi}^{\prime} = \tilde{\Pi}^{\prime} = 0$ at $\Bar{A} = 0$. Additionally, integrating Eq. \eqref{eq:lapseint} along $A$ at each time slice yields the expression for the lapse function:
\begin{equation}\label{eq:lapse}
    e^{\phi} = \tilde{\rho}^{- w_{\rm pf}/(1+w_{\rm pf})}_{\rm pf},
\end{equation}
where $w_{\rm pf}$ is the equation of state of the perfect fluid. This last expression provides the exact analytical solution to the constraint equation, Eq.~\eqref{eq:lapseint}, and has been implemented in the numerical code.  As the Misner-Sharp formalism aligns with the comoving gauge of the fluid (see Eq.~\eqref{eq:comovil}), an increase in fluid density leads to a decrease in the lapse function. Consequently, the proper time experienced by observers near the forming singularity slows down, effectively preserving time resolution. 

Our numerical approach solves Eqs.~\eqref{eq:final}--\eqref{eq:Ufinal} as a set of simultaneous ordinary differential equations using the method of lines (see Section \ref{Sec:IV}). In this framework, $\phi$ is determined analytically using Eq.~\eqref{eq:lapse} from integrating Eq.~\eqref{eq:lapseint}, while $\rho_{\rm pf}$ is computed via Eq.~\eqref{eq:rhoFinal}. A notable computation challenge arises when computing $\Gamma$ in Eq.~\eqref{eq:Gamma} which exhibits a circular dependency—$\Gamma$ depends on the total mass $m_T$, which itself depends on $\Gamma$. We resolve this interdependence through an iterative procedure at each time step to converge on the correct value of $\Gamma$. At $\Bar{A}\rightarrow\infty$ the system should match the FRW background; however, numerical simulations operate on a finite spatial grid. To ensure consistency at the outer boundary, we impose the Neumann boundary conditions $\tilde{P}_{pf}^\prime(\Bar{A},\xi) = 0$ for the perfect fluid. Since the evolution equations contain no $\tilde{P}^\prime_\varphi$ terms, an outer boundary condition is not required. For the scalar field we set $\tilde{m}_\varphi(0,\xi)=\rho_\varphi(0,\xi)$.

\subsection{Condition for Primordial Black Hole Formation}

\subsubsection{Formation of an apparent horizon}\label{PBHconditions}

With the necessary equations derived, we can now evolve initial density perturbations and determine whether they form PBHs. To identify the formation of a black hole, we need a method that does not require knowledge of the entire global structure of the four-dimensional spacetime. Instead, black holes can be characterized using a quasi-local concept, rather than the global concept of an event horizon. The notion of trapped surfaces, from which null rays cannot expand outwards, provides such a quasi-local characterization, allowing the black hole to be identified as a region containing closed trapped surfaces \citep{Helou_2017}. To detect the formation of a black hole, we search for the emergence of a trapped surface, indicating the onset of gravitational collapse. Specifically, we have to consider the expansion of outgoing radial null geodesics, defined as $\Theta^{\pm} \equiv h^{\mu\nu}\nabla_{\mu}k^{\pm}_\nu$. The induced spacetime metric on a spherical surface $\Sigma$ is $h^{\mu\nu} = g^{\mu\nu} + n^\mu n^\nu - s^\mu s^\nu$, where $n^\nu$ is a timelike unit vector orthogonal to slices of constant $t$, and $s^\nu$ is a spacelike unit vector orthogonal to slices of constant $A$. Together, these define the outgoing null vector $k^{\nu(-)}=(n^{\nu}+s^{\nu})/\sqrt{2}$. In our metric, this is $k^{\nu(-)}=(e^{-\phi},e^{-\lambda/2},0,0)/\sqrt{2}$. For spherically symmetric spacetimes, the expansion $\Theta^{\pm}$ is 
\begin{equation}
    \Theta^{\pm} = \frac{1}{4\pi R^2}k^{\mu(\pm)}\nabla_{\mu}(4\pi R^2).
\end{equation}
In the case that $\Theta^{-}<0$ and $\Theta^{+}=0$, a future trapped horizon is formed, characterized by the fact that the ingoing radial null geodesic $\Theta^+$ is vanishing and the outgoing radial null geodesic $\Theta^-$ is non-vanishing. This defines the apparent horizon, which serves as the condition for identifying a black hole. From our metric, we have
\begin{equation}\label{eq:geoexpansion}
    \Theta^{\pm} = \frac{\sqrt{2}}{R}(U\pm\Gamma).
\end{equation}
In terms of our dimensionless variables, Eq.~\eqref{eq:geoexpansion} becomes
\begin{equation}
    \Theta^{\pm} = \sqrt{2}H\bigg(\tilde{U}\pm\frac{\Bar{\Gamma}}{\Bar{A}\tilde{R}}\bigg).
\end{equation}
 An apparent horizon forms when $\Theta^+=0$, corresponding to the condition $U=-\Gamma$ at the horizon. Conversely, when $\Theta^-=0$, we have $U=\Gamma$, which characterises a past trapping horizon associated with an expanding universe/cosmological horizon moving outwards. Combining both cases, we obtain $\Theta^-\Theta^+ = 4/R^2(U^2 - \Gamma^2)$. Since a black hole forms when $U^2=\Gamma^2$, and using Eq.~\eqref{eq:gam2}, we derive the condition for the formation of an apparent horizon in spherically symmetric spacetimes in terms of our evolution variables:
\begin{equation}\label{eq:sfcondition}
    \frac{2M_{\rm MS}}{R }=\frac{2(m_{\rm T}-I_1)}{R} \geq1.
\end{equation}
The physical interpretation of the above condition exceeding unity is that there exist photons which are trapped by the gravitational potential of the central black hole and cannot escape to infinity. Equation \eqref{eq:sfcondition} relates the local mass content, minus a correction term associated with the total energy flux of the scalar field, within a circumferential radius $R$. It can be interpreted as a modified version of the apparent horizon condition for a Schwarzschild black hole.
%\JCH{Hopp conjecture? Is there actually a Schwarzschild condition?} \lp{I think the Hopp conjecture is a generalization to the Schwarzschild condition for no spherical collapses, no? And our Eq. (33) is practically a Schwarzschild-like criteria, but with a correction term that is given by the flux term of the scalar field.}. 
The additional scalar field term indicates that the field’s flux influences the energy distribution across both spatial and temporal coordinates, thereby modifying the location of the apparent horizon with respect to the single perfect fluid case. Notably, this condition depends on the total Misner-Sharp mass of the scalar-fluid system. Expressing this condition in terms of tilde variables yields
\begin{equation}\label{eq:AHcon2}    e^{2(\alpha-1)\xi}\Bar{A}^2\tilde{R}^2\tilde{m}_{\rm T} - \frac{3e^{(2\alpha-3)\xi}}{\Bar{A}\tilde{R}}\mathcal{I}_1\geq 1.
\end{equation}
Therefore, for a PBH to form, the conditions that $\tilde{U}<0$ (since $\Gamma$ is positive) and Eq.~\eqref{eq:AHcon2} must be satisfied simultaneously.% As an example, in Fig.~\ref{fig:ExpansionRadiation}, we plot the condition in Eq.~\eqref{eq:AHcon2} alongside the time slicing of $\Theta^{\pm}$ for the case of radiation fluid and a scalar field under a quartic potential (see Eq.~\eqref{eq:quadratic}) collapsing into a PBH. 
%\begin{figure}[H]
%    \centering
%    \includegraphics[width=1\linewidth]{ExpansionParameterRadiation.png}
%    \caption{\footnotesize{\textbf{Top-panel}:\lp{change this plot. Include also the plot for an scalar field (see main text).} $2(m_{T}-I_1)/R$ as a function of comoving radius $\Bar{A}$ once the trapped surface has formed for the cases where a radiation fluid or a scalar field collapse into a PBH. We have also included a case in which a radiation perturbation does not form a PBH but instead dissipates. The dashed red line corresponds to $2(m_{T}-I_1)/R = 1$. \textbf{Bottom-panel}: Time slice of $\Theta^\pm$ at the same time as the above panel. The vertical dashed black lines correspond to where $\Theta^\pm=0$}}
%    \label{fig:ExpansionRadiation}
%\end{figure}

\subsubsection{Maximum of the compaction function}

Another approach to determining whether a PBH forms is through the compaction function (originally introduced in \citep{Shibata_1999}), which can be expressed in terms of the Misner-Sharp variables as:
\begin{equation}
    \mathcal{C}(A) = 2\frac{\delta m_T(A)}{R(A)}, 
\end{equation}
where $\delta m_{\rm T}(A) = m_{\rm T}(A)-m_{\rm T,b}(A)$ is the mass of the overdensity. A PBH is considered to be close to formation when the maximum value of the compaction function, $\mathcal{C}_{\rm max}$, exceeds unity \citep{Escriva:2019nsa}. Equivalently, in terms of our rescaled (tilde) variables, this condition can be written as:
\begin{equation}
    %R[k,:]**2*(M_t[k,:]-M_t[k,-1])*x**2*np.exp(2*(alp-1)*t[k])
    \mathcal{C}_{\rm max} \equiv \tilde A_{\rm max}^2\tilde R(\tilde A_{\rm max})^2\left[\tilde m_{\rm T}(\tilde A_{\rm max}) - \tilde m_{\rm T,b}(\tilde A_{\rm max}) \right]e^{2(\alpha-1)\xi}\geq 1.
\end{equation}
It is worth noting that this condition differs slightly from the criterion given in Eq.~\eqref{eq:AHcon2}. In this article, we will employ both inequalities as complementary conditions for PBH formation  as described below.

\subsection{Initial Conditions}\label{IC}
We require the initial conditions to be cosmologically consistent with being generated from inflationary perturbations. In particular, the initial perturbations should consist solely of the growing component on scales larger than the cosmological horizon. This requirement reflects the fact that the decaying component would have vanished between the end of inflation and the beginning of our numerical evolution. If the initial perturbations are set at or near the horizon scale or at an insufficient size relative to the horizon scale, then the collapse critical value, $\delta_c$, measured at the cosmological horizon crossing time, may include a substantial decaying component that does not contribute to the PBH formation. To avoid this, the initial data for all relevant variables should be specified on a surface of constant $\xi$ and be comprised of the growing mode only. 

The method for constructing initial conditions that selects the growing component in the linear regime was first developed in \citep{Shibata_1999}. This formalism was clarified and further developed for the perfect fluid dominated case by Bloomfield \textit{et al} \citep{Bloomfield:2015}. In this work, we adopt Bloomfield's formalism in deriving superhorizon initial conditions and extend this description into the scalar field dominated case. In what follows, we assume the simple potential given in Eq.~\eqref{eq:quadratic}. 

\subsubsection{Initial Conditions for a Perfect Fluid Scenario}
In this section, we briefly review the formalism described in \citep{Bloomfield:2015} and outline their procedure for a perfect fluid. Our goal is to determine expressions that relate the dynamical variables in a way that selects only the growing component in the initial data. To begin, we linearise the equations of motion by expressing the variables as
\begin{equation}\label{eq:linearisedX}
    \tilde{X} = 1 + \epsilon\delta_X,
\end{equation}
where $X$ represents the quantities $m_{\rm pf}$, $U$, $R$, $\rho_{\rm pf}$ and $\phi$ and $\epsilon$ is an order counting parameter. The linearised equations of motion are then obtained by substituting the corresponding linearised variable into Eqs.~\eqref{eq:final}, (refer to Eqs.~(68) in \citep{Bloomfield:2015}). These equations can then be manipulated to yield a single second-order PDE:
\begin{equation}\label{eq:dmPDE}
    \begin{split}
    &\partial^2_\xi\delta_{m_{\rm pf}} - (3 - 5\alpha)\partial_\xi\delta_{m_{\rm pf}} + [(1-2\alpha)3w - 1]\\&\times~\alpha\delta_{m_{\rm pf}} = w\alpha^2e^{2(1-\alpha)\xi}\bigg(\frac{4\delta^\prime_{m_{\rm pf}}}{\Bar{A}} + \delta^{\prime\prime}_{m_{\rm pf}}\bigg).
    \end{split}
\end{equation}
If we momentarily neglect the RHS, the equation reduces to an ODE, allowing for a solution of the form $\delta_{m_{\rm pf}}=C\text{exp}(\beta\xi)$. The corresponding characteristic equation yields the solutions 
\begin{equation}
    \beta_1 = \frac{w-1}{w+1}~~~~\text{and}~~~~\beta_2=2(1-\alpha).
\end{equation}
For $-1<w<1$, the first solution is a decaying mode, while for $w>-1/3$, the second solution is a growing mode. For the perfect fluid scenario, and since we picture a non-accelerating expansion, we assume $w>-1/3$, and retain only the growing solution, which functions as the principal initial condition for the data
\begin{equation}\label{eq:dmsol}
    \delta_{m_{\rm pf}}(\Bar{A},\xi) = \delta_{{m0}_{\rm pf}}e^{2(1-\alpha)\xi}.
\end{equation}
Letting $\delta_{m0_{\rm pf}}=\delta_{m_{\rm pf}}(\Bar{A},\xi=0)=\delta_{m_{\rm pf}}(\Bar{A},t=t_0)$. With this expression for $\delta_{m_{\rm pf}}$, the remaining perturbative first order solutions can be derived. Returning to the terms omitted on the RHS of Eq.~\eqref{eq:dmPDE}, Bloomsfield \textit{et al} introduce henceforth a  gradient expansion in $\delta^0_{m_{\rm pf}}$. The solution found in Eq.~\eqref{eq:dmsol} is the zeroth order solution of $\delta^0_{m_{\rm pf}}$. Terms containing derivatives are suppressed relative to terms without derivatives by introducing a scaling transformation $\Bar{A}\rightarrow\Bar{A}/\sqrt{\varepsilon}$, where $\varepsilon$ is an order counting parameter for the derivative expansion different from $\epsilon$ denoting perturbative ordering. Performing this scaling transformation on Eq.~\eqref{eq:dmPDE} demotes terms on the RHS to first-order contributions in the derivative expansion. The solution $\delta_{m_{\rm pf}}$ can then be expressed as a series expansion and, therefore, successively higher-order terms become increasingly negligible.

The solutions to the equations of motion can be extended beyond linear order in perturbation theory, as demonstrated in \citep{Bloomfield:2015} for the perfect fluid scenario.  In our case, the numerical code is designed to evolve the equations of motion using second-order initial conditions, where the initial data is fully constructed from the linear mass perturbation $\delta_{m0_{\rm pf}}$. However, for the scalar field-dominated case, we truncate our initial conditions to first-order  as the complexity of the resulting system of equations makes higher-order corrections impractical (see the next section for details).

\subsubsection{Initial Conditions for an scalar field-dominated Scenario}
Following the procedure outlined in the previous section, we now derive approximated initial conditions for an scalar field-dominated scenario, considering both the perturbative and gradient expansions up to first order.\footnote{For an alternative derivation of appropriate initial conditions for an scalar field-dominated universe in the constant mean curvature slicing, see Ref.~\citep{Padilla:2021uof}.} Given that our code is set in a coordinate system that moves with the perfect fluid, we cannot, in general, simply set the fluid contributions to zero. Instead, we consider an extremely diluted, homogeneous, and isotropic fluid, and express the linearized quantities for the scalar field and metric variables as expressed in Eq.~\eqref{eq:linearisedX}. In this scenario, we highlight that following quantities differ from the perfect fluid case:
\begin{subequations}
\begin{equation}
    \tilde{\varphi} = \tilde{\varphi}_{\rm b} + \epsilon\delta_\varphi,
\end{equation}
\begin{equation}
    \tilde{\chi} = \epsilon\delta_\chi,
\end{equation}
\begin{equation}
    \tilde{\Pi} = \tilde{\Pi}_{\rm b} + \epsilon\delta_\Pi.
\end{equation}
\end{subequations}
Substituting these linearised quantities into the equations of motion given in Eq.~\eqref{eq:final} and using the potential Eq.~\eqref{eq:quadratic} we obtain the following relevant first-order perturbative equations:
\begin{widetext}
\begin{subequations}
\begin{eqnarray}
    \partial_\xi\delta_\varphi = \frac{n+1}{3n} e^\xi\delta_\Pi + \delta_\varphi,
\end{eqnarray}
\begin{eqnarray}
    \partial_\xi\delta_\Pi = \delta_\Pi - \frac{n+1}{3n}\Bigg[ 3\delta_\Pi + \bigg(3\delta_U + \Bar{A}\delta^\prime_U + \frac{3}{2}e^{-\xi}\Bar{A}\delta_\chi\tilde{\Pi}_{\rm b}\bigg)\tilde{\Pi}_{\rm b} + e^{(3-2n)\xi} \tilde{\lambda}_n\tilde\varphi_b^{2(n-1)}\delta_\varphi\Bigg],
\end{eqnarray}
\begin{eqnarray}\label{eq:deltaR}
    \partial_\xi\delta_R = \frac{n+1}{3n}\delta_U,
\end{eqnarray}
\begin{eqnarray}\label{eq:deltaU}
    \partial_\xi\delta_U =\left(1-\frac{2(n+1)}{3n}\right)\delta_U -\frac{n+1}{6n}\left( \delta_{m_\varphi} + 3\delta_{P_\varphi} - \frac{3e^{-\xi}}{\Bar{A}^3}\int^{\Bar{A}}_0\Bar{A}^3\delta_\chi\tilde{\Pi}_{\rm b}d\Bar{A}\right),
\end{eqnarray}
\begin{eqnarray}\label{eq:deltarho}
    \delta_{\rho_\varphi} = \tilde\Pi_{\rm b}\delta_\Pi + {\tilde{\lambda}_n}e^{2(1-n)\xi}\tilde\varphi_b^{2n-1}\delta_\varphi,
\end{eqnarray}
\end{subequations}
\end{widetext}
where $\tilde{\lambda}_n\equiv R_H^2\lambda_n(3/8\pi)^{n-1}$ and we have used that $\alpha = (n+1)/(3n)$. 

Note that the absence of $\delta_R$ in all evolution equations, except for \eqref{eq:deltaR}, arises because the initial data for $\tilde{R}$ is a gauge choice. The physical initial data is $\tilde{m}_\varphi(R)$ (or the scalar field variables) and $\tilde{U}(R)$.

Due to the complexity of the system, we cannot reduce the equations into a single second-order ODE and solve directly for one of the variables. Instead, we will utilize different simplifications to derive appropriate initial conditions. Specifically, we start by adopting the approach of generating appropriate cosmological initial conditions based on an initial curvature profile \citep{Polnarev_2007, Polnarev_2012, Nakama_2014}, which was claimed to pick correctly the growing mode. The curvature profile $K(A,t)$ is defined by writing 
\begin{eqnarray}
    e^{\lambda/2}(A,t) = \frac{R^\prime(A,t)}{\sqrt{1 - K(A,t)A^2}}.
\end{eqnarray}
For a curved FRW universe, we have the relation:
\begin{eqnarray}\label{eq:K}
    \Gamma^2 = 1 - KA^2,
\end{eqnarray}
which shows the connection between the profile $K$ and $\Gamma$, which is an invariantly defined quantity. We can then relate $K$ to our Misner-Sharp variables by matching Eq.~\eqref{eq:K} to Eq.~\eqref{eq:Gsquared} and defining $\tilde{K}\equiv K R_H^2$. This allows us to characterise the cosmological perturbations in terms of the curvature perturbation $K$ as
\begin{equation}\label{eq:CurvProf}
    \begin{split}
    \tilde{K} = e^{\frac{2(1-2n)}{3n}\xi}&(\delta_{m_{\varphi}} - 2\delta_U) - \frac{3\Pi_{\rm b} e^{\frac{2-7n}{3n}\xi}}{\Bar{A}^3}\int^{\Bar{A}}_0 \Bar{A}^3\delta_\chi d\Bar{A},
    \end{split}
\end{equation}
where we have considered the simplification $\delta_{m_{\rm T}} \simeq \delta_{m_{\varphi}}$. To construct the initial data, we assume that at superhorizon scales, the curvature profile is a time-independent quantity, $K(A,t) = K(A)$. Under this assumption, we expect that both $\delta_U$ and $\delta_{m_{\varphi}}$ have a time-dependence of $\propto \text{exp}[2(2n-1)\xi/3n]$. Notably, this is the same time dependence shown in Eq.~\eqref{eq:dmsol} for a perfect fluid with equation of state $\omega = (n-1)/(n+1)$. 

The energy density in the scalar field can be expressed as:
\begin{equation}
    \begin{split}
    \tilde{\rho}_{ \varphi} &= \frac{\tilde{\Pi}^2}{2} + \frac{\Bar{\Gamma}^2\tilde{\chi}^2}{2(\Bar{A}\tilde{R})^{\prime2}}e^{-2\xi} + \frac{\tilde{\lambda}_n}{2n}\tilde{\varphi}^{2n} \\
    &=\tilde{m}_{\varphi} + \frac{\Bar{A}\tilde{R}}{3(\Bar{A}\tilde{R})^\prime}\tilde{m}^\prime_{\varphi},
    \end{split}
\end{equation}
which when linearised yields Eq.~\eqref{eq:deltarho}
\begin{equation}
    \tilde{\Pi}_{\rm b}\delta_\Pi + {\tilde{\lambda}_n}e^{2(1-n)\xi}\tilde\varphi_b^{2n-1}\delta_\varphi = \delta_{m_{\varphi}} + \frac{\Bar{A}}{3}\delta^\prime_{m_{\varphi}}\propto e^{2(2n-1)\xi/3n}.
\end{equation}
We simplify our description by assuming that at $\xi = 0$ (the initial time in our numerical simulations), we have $\tilde\varphi_{\rm b0} = 0$ and then $\tilde \Pi_{\rm b0}^{2}/2 = 1$. Our major simplification will be to assume that at time $\xi = 0$, ${\tilde \varphi_0} = \tilde\chi_0 = 0$, so all the information of the initial profile that we will use is contained in the kinetic term of the scalar field. With these assumptions, the above equation reduces to
\begin{equation}
    \delta_{\Pi 0} = \frac{1}{\sqrt{2}}\bigg(\delta_{m0_{\varphi}} + \frac{\bar A}{3}\delta^\prime_{m0_{\varphi}}\bigg),
\end{equation}
where $\delta_{m0_\varphi}$ is an initial mass perturbation of the scalar field.

We now proceed to obtain an initial condition for $\delta_U$. From Eq.~\eqref{eq:deltaU} and using the assumption that $\delta_U\propto \text{exp}[2(2n-1)\xi/3n]$ and $\delta_{\chi 0} = 0$, we obtain:
\begin{equation}
    \delta_{U0} = -\frac{n+1}{6n}\left( \delta_{m0_\varphi} + 3\delta_{P0_\varphi} \right).
\end{equation}
Next, we recognise that, for our initial condition $\delta_{P0_\varphi} \simeq \delta_{\rho 0_\varphi} \simeq \tilde{\Pi}_{\rm 0b}\delta_{\Pi 0}$, we obtain
\begin{equation}
    \delta_{U0} = -\frac{n+1}{6n}\left( 4\delta_{m0_\varphi} + A\delta^\prime_{m0_{\varphi}} \right)
\end{equation}
Observe that, substituting this expression into Eq.~\eqref{eq:CurvProf} and assuming again that $\delta_{\chi 0} = 0$, we can immediately obtain a relation between $\delta_{m0_\varphi}$ and the initial curvature perturbation $\tilde K$. This means that we can use either $\delta_{m0_\varphi}$ or $\tilde K$ as our initial profile to evolve numerically. However, given that all our quantities are given in terms of $\delta_{m0_\varphi}$, we will use this profile in our numerical simulations, noting the direct relationship with the curvature perturbation.

To conclude this section, we need to give an initial condition for $\delta_R$. Taking Eq.~\eqref{eq:deltaR} and using once again that $\delta_U\propto \text{exp}[2(2n-1)\xi/3n]$, we finally obtain:
\begin{equation}
    \delta_{R0} = -\frac{(n+1)^2}{12n(2n-1)}\left( 4\delta_{m0_\varphi} + A\delta^\prime_{m0_{\varphi}} \right).
\end{equation}

\section{Numerical implementation}\label{Sec:IV}

The numerical simulations are performed using our newly developed numerical code based on the prescription of the hydrodynamical Misner-Sharp evolution equations presented in this work. The perfect fluid Misner-Sharp equations are modified by the addition of a scalar field contribution. That is, to determine the growth of the inhomogeneity after horizon re-entry, we solve the Einstein equations with both a perfect fluid and scalar field component. The initial conditions are discretised across the computational grid, where we utilise finite difference methods to calculate gradient terms throughout this grid. The system's time evolution is performed using the method of lines, where the dynamical variables ($\varphi$, $\Pi$, $U$, $m$, $R$) are propagated forward in time using a fourth-order Runge-Kutta integrator.
%Both the space and time coordinates are discretised following the method of lines. 
The evolution forward in time is achieved through time stepping $\xi\rightarrow\xi + \Delta\xi$, where at each time step the evolution and constraint equations are computed. We numerically integrate the evolution equations along the logarithmic time coordinate $\xi$  using $4^{th}$ order accurate Runge-Kutta method. The spatial derivatives are performed using $\mathcal{O}(\Delta\Bar{A}^4)$ finite difference methods where we take advantage of the symmetries of the variables across the inner boundary. In this way, we impose reflective boundary conditions, implicitly allowing for the first derivative of the variables to vanish at the boundary. At the outer boundary, explicit Neumann boundary conditions are imposed for any variable $f$,
\begin{equation}
    \frac{\partial f}{\partial \Bar{A}} = 0.
\end{equation}
Since the strength and variability of the field variables are significant only at a small region, it is crucial to have a system capable of capturing the essential features of the dynamics. For that reason, our code contains regridding procedures to adaptively increase the number of grid points by interpolating data to achieve higher resolution as the spacetime imminently collapse to form a PBH. For finite difference methods, large numerical instabilities can arise from the appearance of high frequency spurious modes and interpolation associated errors. We stabilise the method by implementing a $N=3$ Kreiss-Oliger \citep{Kreiss} dissipation to damp out the formation of high frequency modes. For all evolution variables $f \in \{U,m,R,\varphi,\chi,\Pi\}$, the evolution equations are modified as follows
\begin{equation}
    \begin{split}
    \partial_\xi f \rightarrow& \partial_\xi f \\&+ \frac{\sigma}{64\Delta\Bar{A}}(f_{m+3} - 6f_{m+2} + 15f_{m+1} \\& - 20f_m +15f_{m-1} - 6f_{m-2} + f_{m-3}),
    \end{split}
\end{equation}
where $m \pm n$ labels the grid point $m$, $n$ the total offset from $m$ and $\sigma$ is an adjustable dissipation parameter usually of the order $\mathcal{O} (10^{-2})$.

In all the simulations, we fix the value $a(t_0)=1$, $\xi_0 = 0$ defined by $t=t_0e^\xi$ and $w=1/3$ or $0$ for the perfect fluid or scalar field cases under a quartic and quadratic potential (see appendix \ref{app:background}). Automatically, we obtain $R_{H} = t_0/\alpha$ and $H_0 = \alpha/t_0$. When simulating the scalar field collapse, we adopt a radiation-like reference fluid with an equation of state parameter $w_{pf}=1/3$, which aligns with most cosmological scenarios of interest. To ensure that the dynamics of the perfect fluid do not significantly influence the evolution, we consider a subdominant radiation component, maintaining an energy density ratio of $\tilde{\rho}_\varphi/\tilde{\rho}_{pf} \simeq \mathcal{O}(10^8)$. This choice allows us to focus on the primary effects of the collapsing scalar field while preserving consistency with a radiation-dominated background.

As for the initial profile of the overdensity, we considered a simple Gaussian profile for $\delta_{m_T}$, i.e.:
\begin{equation}
    \delta_{m_T} = \kappa e^{-A^2/2\sigma^2},
\end{equation}
where, for simplicity, we used in all our simulations the value $\sigma = 2R_H$,\footnote{Note that this value is just slightly larger than the cosmological horizon. We can make this simplification due to the fact that in the previous section we showed that our initial conditions correctly select only the growing mode of the solutions.} leaving $\kappa$ as the only free parameter in our simulations.

Finally, if the fluctuation evolves towards PBH formation, it is crucial to identify the emergence of an apparent horizon. To achieve this, we systematically check whether the two conditions discussed in \ref{PBHconditions} are simultaneously satisfied across the computational grid. Additionally, we impose the requirement that the compaction function exceeds unity. The fulfillment of these criteria is interpreted as the formation of a PBH. We define the precise time of PBH formation, denoted as $\xi_{AP}$, as the first time step at which all three conditions are simultaneously satisfied at any point in our computational grid. This allows us to systematically compare formation times across different initial configurations. 

\section{Testing the code}\label{Sec:V}

\subsection{The background universe for perfect fluid- and scalar field-dominated universes}
To validate the accuracy of our code, we compute the relative errors of different variables with respect to the background analytical FRW solutions. We define $\delta X_i = X(\Bar{A}_i) - X_b(\Bar{A}_i)$, where $X$ are the Misner-Sharp variables at the grid points. To test whether our code is in agreement with the FRW solution, we compute the $L^2$ norm at each time step 
\begin{equation}
    ||\delta X||_2 = \frac{1}{N}\sqrt{\sum_i|\delta X_i|^2}.
\end{equation}
For the perfect-fluid dominated scenario, the $L^2$ norm computed gives an exact agreement between the numerical and analytical solutions for the background FRW variables by design.

\begin{figure}[h]
    \centering
    \includegraphics[width=1\linewidth]{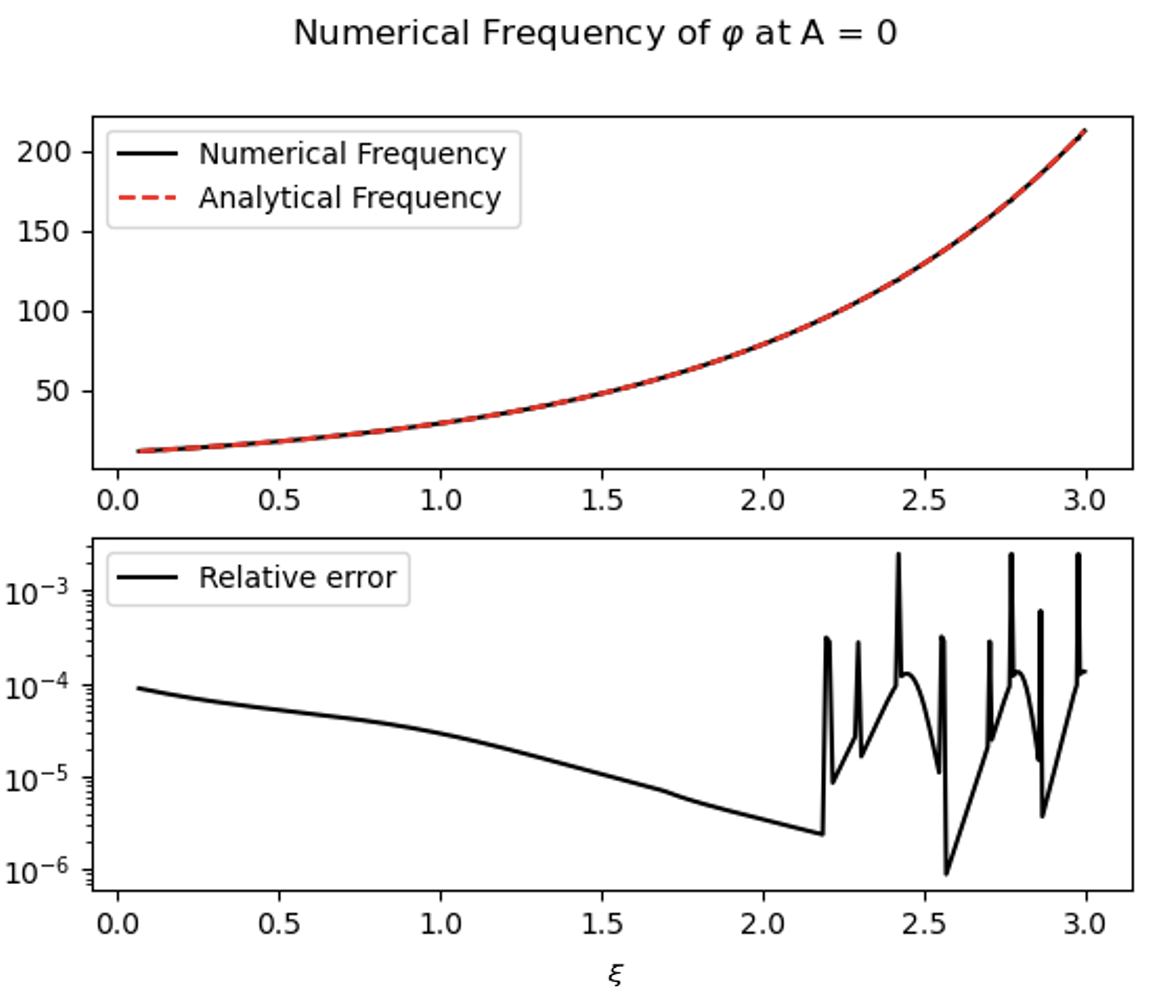}
    \caption{\footnotesize{\textbf{Top-panel}: Numerical and analytical dimensionless frequency of the background universe in a quadratic dominated scenario. \textbf{Bottom-panel:} Relative error between the analytical and the numerical solutions.}}
    \label{fig:numfreq}
\end{figure}

In the case of the scalar field-dominated scenario, we limit our comparison to the case with a quadratic potential, as analytical solutions exist for this case. In particular, if the oscillation of $\varphi$ is sufficiently undamped, which is the case if $2\mu\gg3H$, then the friction term in the Klein-Gordon equation can be neglected. Therefore, the background dynamics of the scalar field evolves according to (see appendix \ref{app:background}):
\begin{equation}
    \tilde \varphi_{\rm b} = \tilde\varphi_{\rm 0,b}\cos\left(\frac{2}{3}\tilde\mu e^\xi+\delta_0\right).
\end{equation}
We see that $\varphi$ oscillates with angular frequency that grows exponentially in $\xi$. The background dynamics of the scalar field are successfully replicated numerically by our code as presented in Fig \ref{fig:numfreq}, where the relative error between the numerical and analytical solutions remains below $10^{-3}$. 
\begin{comment}
The resulting form of the background pressure expressed in cosmic time $t$ is
\begin{equation}
    P = \frac{\phi_0^2\mu^2}{2}\bigg(sin^2(\mu t) - cos^2(\mu t)\bigg).
\end{equation}
This form is identical to the one used in \citep{deJong:2021bbo}
\end{comment}

\subsection{Hamiltonian Constraints}
{A key diagnostic for the reliability of our simulations is the consistency with the Hamiltonian constraint, Eq.\eqref{eq:Gamma}. Since we do not evolve $\Bar{\Gamma}^2$ using an independent evolution equation, but rather compute it iteratively at each time step from the constraint itself and then use it within the evolution system, we cannot directly compare it to a dynamically evolved quantity. Instead, to assess whether the constraint is being consistently satisfied, we differentiate Eq.\eqref{eq:Gamma} with respect to the spatial coordinate $\Bar{A}$ and monitor whether the resulting relation holds throughout the simulation. This provides an alternative but equivalent check up of the consistency of our method against the Hamiltonian constraint. Our consistency test evaluates the following expression:}
\begin{eqnarray}\label{eq:HC}
   &&{\mathcal{H}} = \bar A^2 \tilde R^2 \tilde U\left[{(\bar A\tilde{R}\tilde{U})^{\prime} + \frac{3}{2}e^{-\xi}\Bar{A}\tilde{R}\tilde{\chi}\tilde{\Pi}}\right] \\
    && - {(\bar A\tilde R)^{\prime}}\left[\frac{\bar A\tilde R\bar \Gamma'}{e^{\lambda/2-\xi}}+\frac{3}{2}\bar A^2\tilde R^2\tilde \rho_{\rm T} + \frac{\bar \Gamma^2 -(\bar A\tilde R\tilde U)^2 - e^{2(1-\alpha)\xi}}{2}\right].\nonumber 
\end{eqnarray}
{This difference should be close to zero in each timestep. In order to keep control over the error, we define a normalized version of $\mathcal{H}$ and consider the simulation to be numerically consistent if the magnitude of the Hamiltonian constraint remains below $10^{-2}$ throughout the evolution.}

\subsection{Primordial black hole formation for a radiation-dominated universe}

\begin{figure}
    \centering
    \includegraphics[width=9.5cm, height=5.5cm]{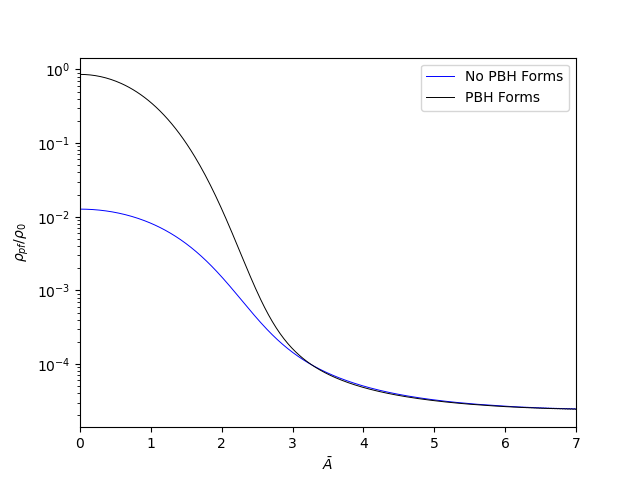}
    \caption{\footnotesize{Density profile $\rho_{\rm pf}/\rho_0$ as a function of comoving radius $\Bar{A}$. The initial conditions for the black line were taken above the threshold for collapse, whereas the initial conditions for the blue line were taken below the threshold for collapse.}}\label{fig:peakrhodimless}
    %\kappa=0.175}
\end{figure}

\begin{figure}
    \centering
    \includegraphics[width=9cm, height=5cm]{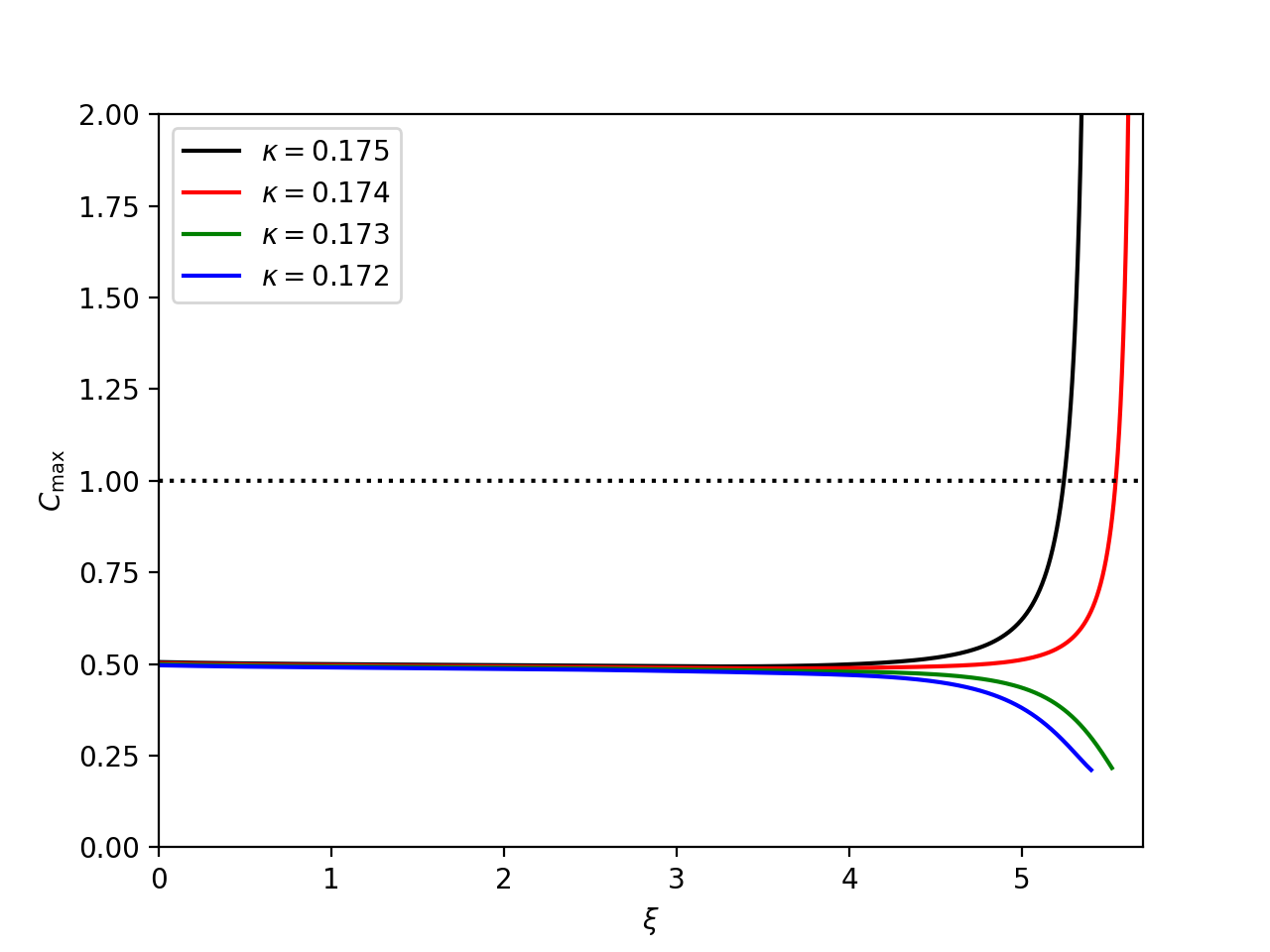}
    \includegraphics[width = 9cm, height=5cm]{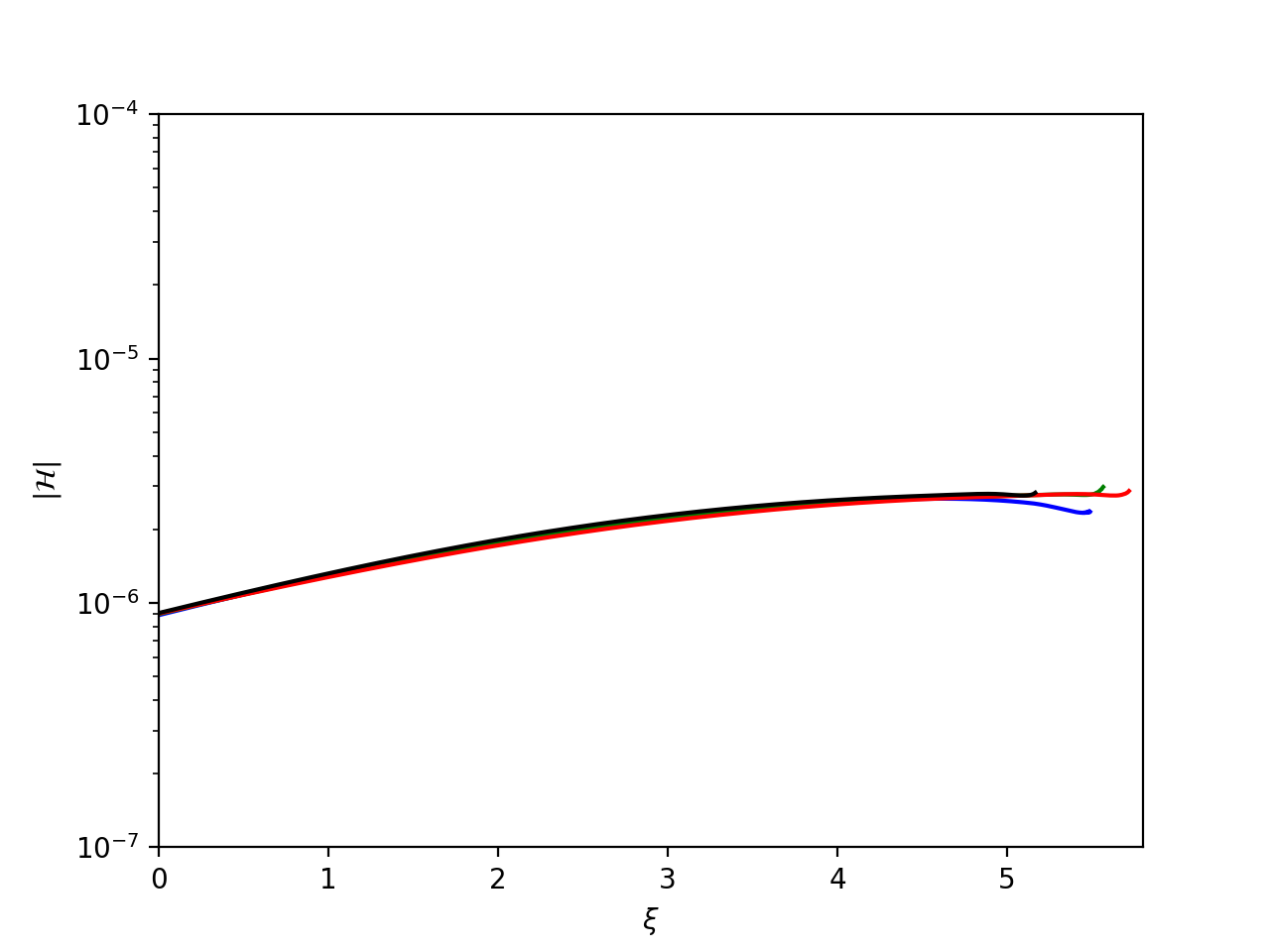}
    \includegraphics[width=8.5cm, height=5cm]{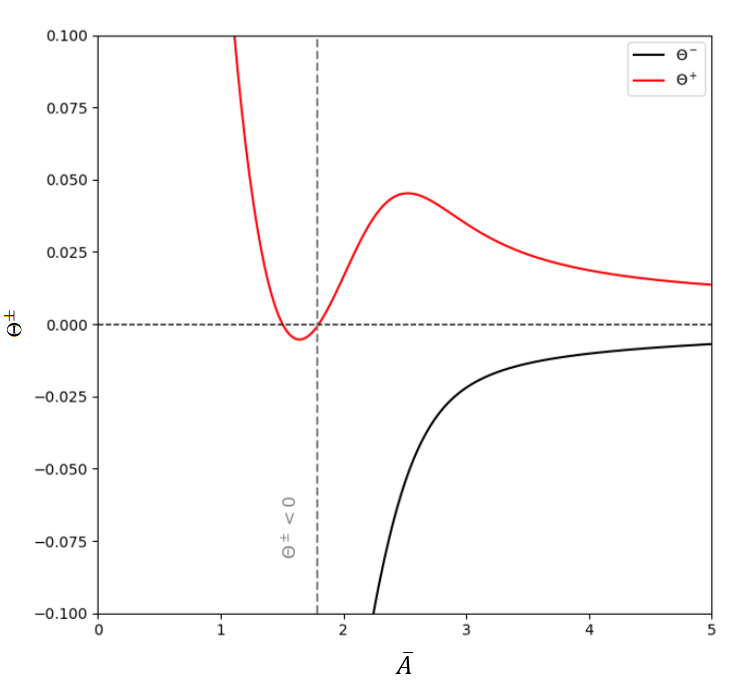}
    \caption{\footnotesize{\textbf{Top-panel:} Evolution of the maximum of the compaction function for different initial conditions for a radiation-dominated universe. \textbf{Middle panel:} {Mean value of the Hamiltonian constraint violation for each case shown in the top-panel.} \textbf{Bottom panel}: Final time slicing, with a vertical line marking the location where the outgoing radial null geodesic vanishes, satisfying the condition $\Theta^{\pm}<0$ for a future trapped horizon. The condition that $\Theta^+=0$ and $\Theta^- < 0$ identifies the outermost trapped surface; the apparent horizon}}\label{Fig:2ddensity}
\end{figure}

As a validation test to confirm that our code accurately reproduces previous results for a radiation-dominated universe, this section focuses on replicating a key result from \citep{Bloomfield:2015}. The authors demonstrated that an initial condition with $\kappa = 0.175$ leads to the collapse of the perturbation into a PBH, while $\kappa = 0.173$ results in the dissipation of the system due to radiation-induced pressure. Fig.~\ref{fig:peakrhodimless} shows the peak energy density for both cases prior to either PBH formation or perturbation dissipation. Our results closely match Figure 12 in Ref.~\citep{Bloomfield:2015}. Additionally, in the upper panel of Fig.~\ref{Fig:2ddensity}, we present the evolution of the maximum of the compaction function for different initial $\kappa$ values, clearly illustrating the transition between collapsing and non-collapsing perturbations. {The middle panel presents the corresponding mean value of the Hamiltonian constraint expression $\mathcal{H}$, demonstrating the numerical consistency of our simulations in all cases.} Finally, the bottom panel of Fig.~\ref{Fig:2ddensity} further depicts the expansion of outgoing radial null geodesics for $\kappa = 0.175$, where the formation of an apparent horizon is evident. 

A detailed analysis of Fig.~\ref{Fig:2ddensity} indicates that the threshold value for PBH formation in our simulations is $\kappa = 0.1735$, corresponding to a maximum compaction function value of $\mathcal{C}_{\rm th} = 0.50$, evaluated at the time of horizon crossing, $\xi_{\rm HC} = 1.25$. This result is consistent with previous findings, where the threshold $\mathcal{C}_{\rm th}$ for PBH formation has been reported as $\mathcal{C}_{\rm th} = 0.5$ for a Gaussian curvature perturbation \citep{Musco:2018rwt}.  

To further complement this section, we can also express the threshold value for PBH formation in terms of the average fractional mass excess, $\delta$, defined as:  
\begin{equation}
    \delta = \frac{m(\bar{A})}{m_b(\bar{A})} - 1,
\end{equation}  
or in terms of our tilde variables:  
\begin{equation}
    \delta = \tilde{m}(\bar{A}) - 1.
\end{equation}  
Once evaluating the above expression at the time of horizon crossing and averaged over the horizon size $A_H$, we obtained $\delta_{\rm th} = 0.45$.


%In the case of the $\phi^2$ potential, as the scalar field perturbation evolves towards gravitational collapse, the spatial extent becomes increasingly confined towards the inner boundary, especially for smaller initial amplitudes. To adequately resolve the growing gradients and the steepening of the metric functions, we perform a series of regridding operations, progressively increasing the spatial resolution from an initial $N=800$ up to $N=10,000$. However, we note that even with the regridding method and high resolution, the simulation eventually reaches a point at which the Hamiltonian constraint violation exceeds $\sim\times10^{-2}$, which we are considering the limit on numerical reliability. This increase is not due to a fundamental issue with the dynamics, but rather signals the limit of numerical stability under in our current simulation. Within this regime, the dynamics are robustly captured, and we can confidently identify black hole formation. Given these limitations, we choose not to explore initial data with perturbation amplitudes smaller than our current results reported in this work.

%\begin{figure}[H]
%    \centering
%    \includegraphics[width=1\linewidth]{rhoRaditionfinalTimeSlice.png}
%    \caption{\footnotesize{Energy density $\tilde{\rho}_R$ map taken at the logarithmic time-slice at which an aparent-horizon forms as a function of the comoving radius $\Bar{A}=\Bar{x}=\Bar{y}$ in spherical symmetry. The dotted white circle indicates the apparent horizon.}}\label{Fig:density}
%\end{figure}
%kappa = 0.175

\section{Primordial black hole formation in scalar field dominated systems}\label{Sec:VI}

\subsection{The case of a $\phi^4$ potential}

\begin{figure}
    \centering
    \includegraphics[width=8.5cm,height=5cm]{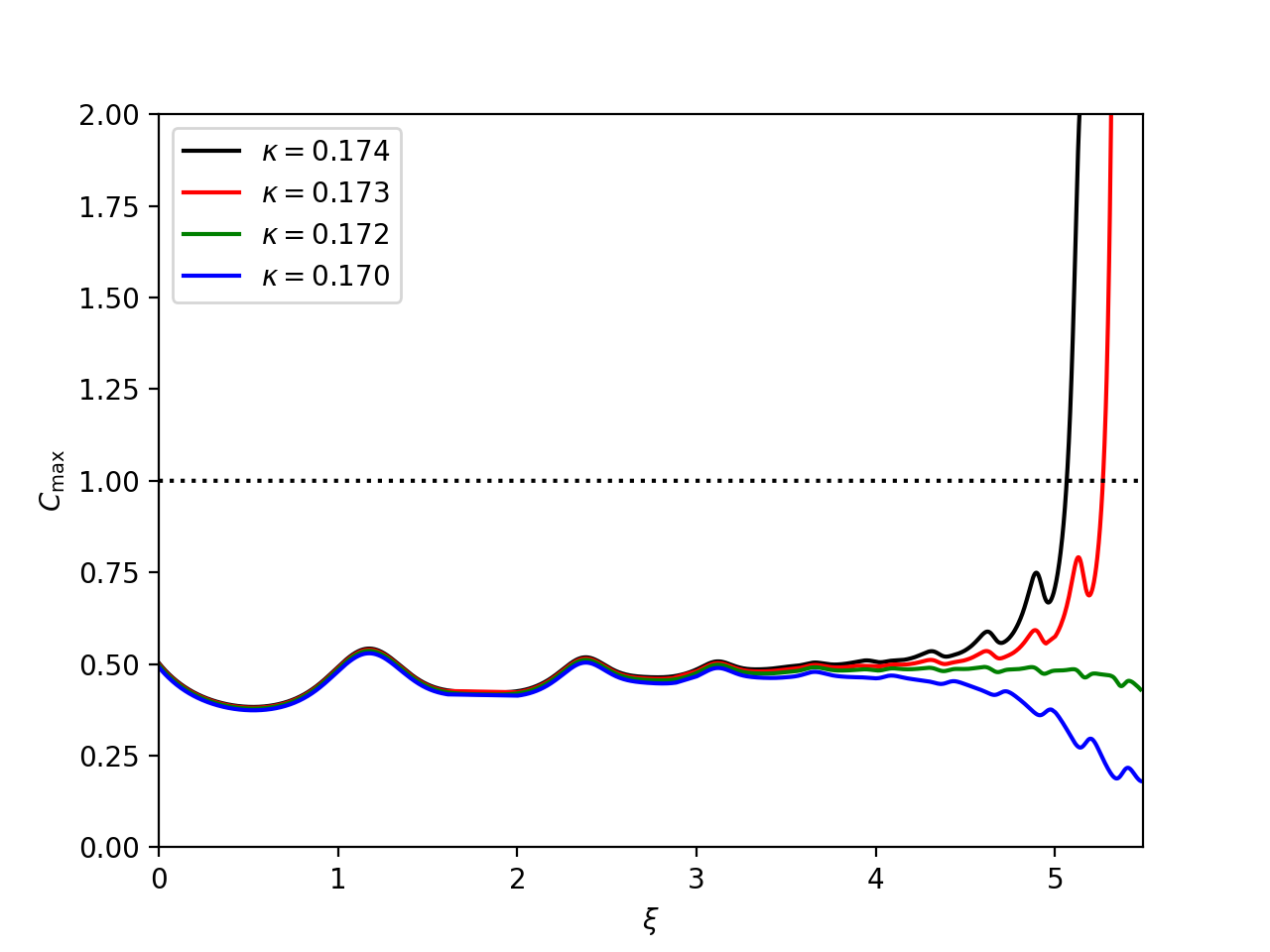}
    \includegraphics[width = 8.5cm,height=5cm]{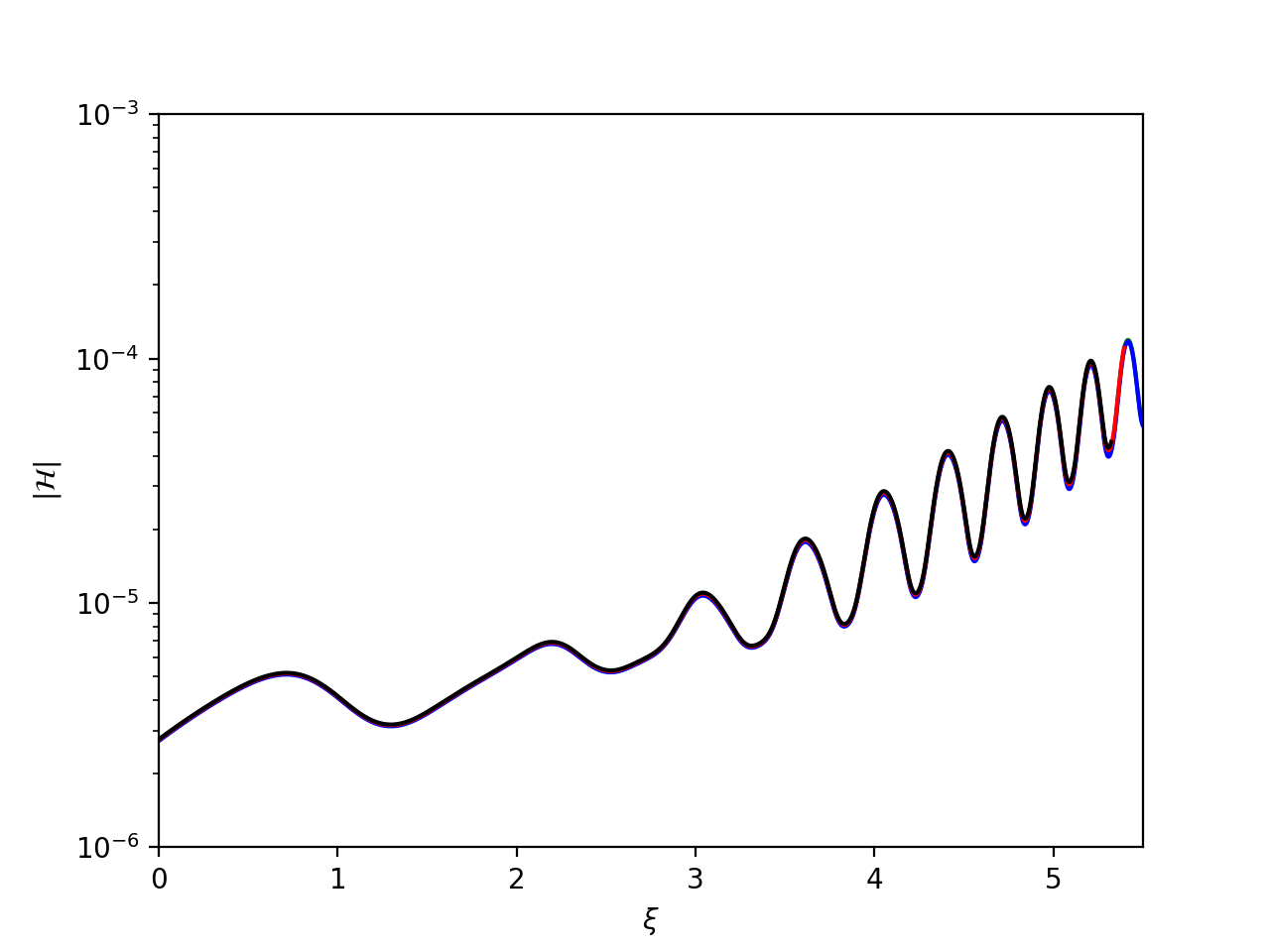}
    \includegraphics[width=8.5cm, height=5cm]{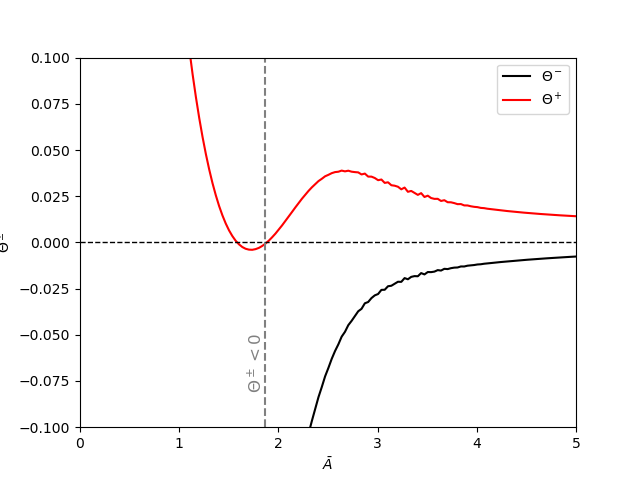}
    \caption{\footnotesize{Same as Fig.~\ref{Fig:2ddensity} but for the quartic/self-interacting scenario.}}
    \label{fig:compaction_phi4}
\end{figure}

In this section, we analyse the case of a scalar field with a quartic potential as the dominant component. Given its expected similarity to the radiation-dominated scenario (see Eqs.~\eqref{eq:quadratic} and \eqref{eq:omega}), we anticipate that the threshold for PBH formation in this case will be close to that of radiation-dominated scenario, specifically with $\kappa_{\rm th} = 0.1735$ and $\mathcal{C}_{\rm th} = 0.50$, or $\delta_{\rm th} = 0.45$. Considering this expectation, we examine the evolution of different initial conditions with various values of $\kappa$, keeping them close to the threshold for PBH formation in the radiation-dominated case. Figure~\ref{fig:compaction_phi4} reproduces the same quantities as Fig.~\ref{Fig:2ddensity}, but for the quartic potential scenario. In all our simulations, we set the self-interaction parameter to $\tilde{\lambda} = 10$. As seen in the figure, the threshold value for PBH formation in this case is found to be $\kappa_{\rm th} = 0.1725$, differing only slightly from the radiation-dominated value.

A clear difference arises when comparing the evolution of the maximum of the compaction function between the radiation-dominated and scalar field-dominated cases. In the latter, oscillations in the compaction function are observed, whereas in the former, they are absent. These oscillations introduce a limitation in our results, leading to an ambiguity in determining the precise threshold value for PBH formation in the quartic potential scenario. To mitigate this issue, we compute the mean value around which the compaction function oscillates and evaluate it at the moment of horizon crossing, obtaining the threshold values $\mathcal{C}_{\rm th} = 0.50$ and $\delta_{\rm th} = 0.45$, just as in the radiation dominated universe.

\subsection{The case of a $\phi^2$ potential}

Following the analogy of a perfect fluid, in this case, we expect scalar field perturbations to evolve similarly to dust. That is, any spherically symmetric perturbation, such as those studied here, would be expected to inevitably collapse into a PBH, regardless of the initial amplitude of the perturbation. However, it is well known that in the quadratic case, there is a crucial difference that distinguishes the nonlinear evolution of scalar field perturbations from that of dust. This difference is due to an effective pressure generated by the wave-like nature of the field, which has been widely studied in the Newtonian regime \citep{Dawoodbhoy:2021beb, PhysRevD.97.083519, Kopp:2017hbb}. This pressure has also been interpreted as a consequence of Heisenberg’s uncertainty principle, which opposes gravitational collapse. Instead, it allows the system’s self-gravity to balance, leading to the formation of virialised structures. In this interpretation, such effective pressure is often referred to as quantum pressure \citep{Padilla:2020sjy, Ralegankar:2024zjd}.

In the context of gravitational collapse with asymptotic vacuum conditions, it has been shown that scalar field distributions—both spherical and non-spherical—tend to relax through a gravitational cooling mechanism, forming solitonic structures \citep{PhysRevLett.72.2516}. These solitons are precisely the result of the balance between quantum pressure and gravity. In a cosmological setting, studies in the weak-gravity regime with subhorizon initial conditions have shown that, under such simplifications, the formation of a soliton is an attractor solution for the system. In this context, the solitons are accompanied by an envelope that, on average, resembles a Navarro-Frenk-White profile which asymptotically decays as $r^{-3}$ for large radii \citep{2014NatPh..10..496S}.

\begin{figure}
    \centering
    \includegraphics[width=9.5cm, height=5.5cm]{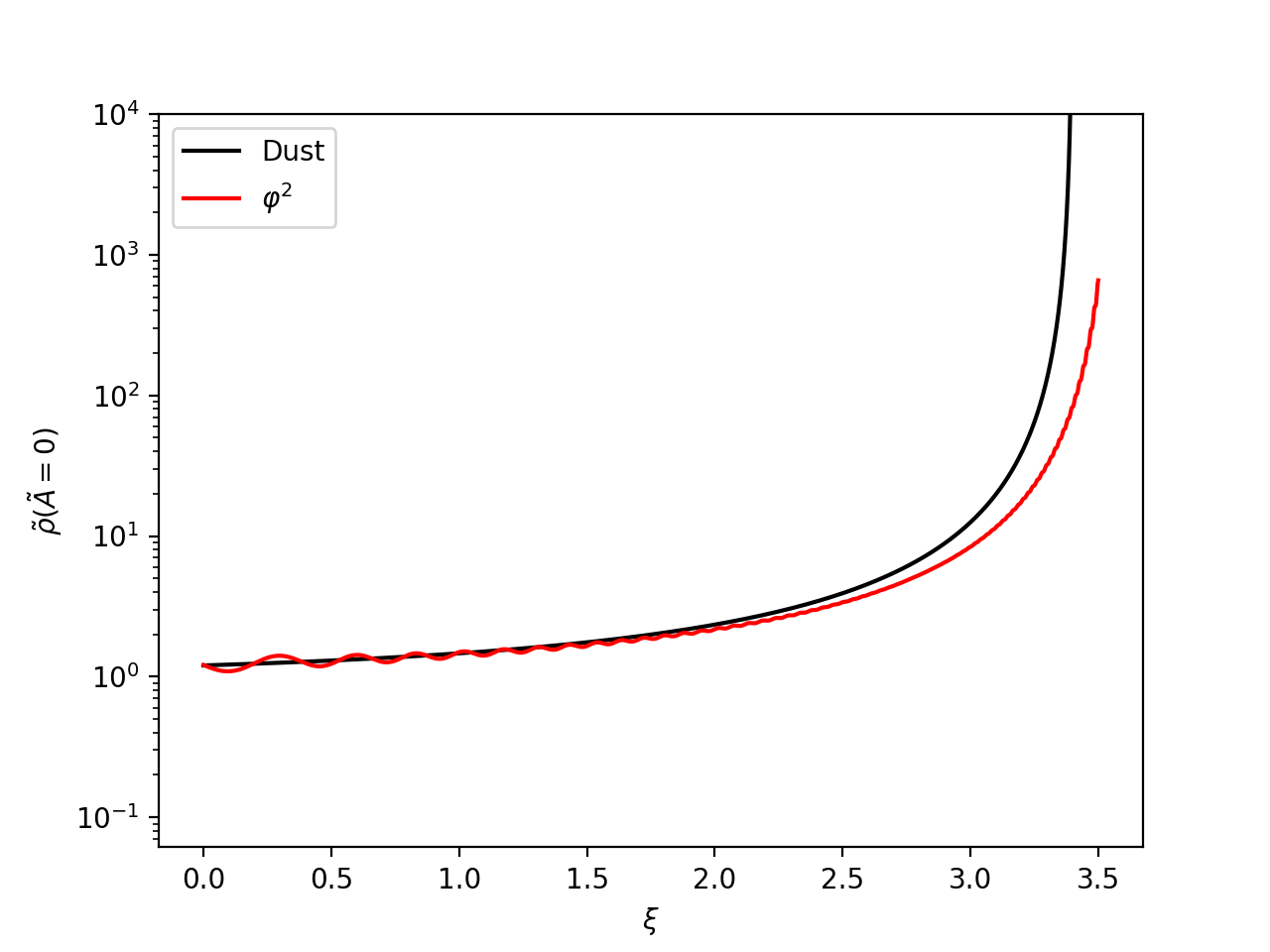}
\includegraphics[width=9.5cm, height=5.5cm]{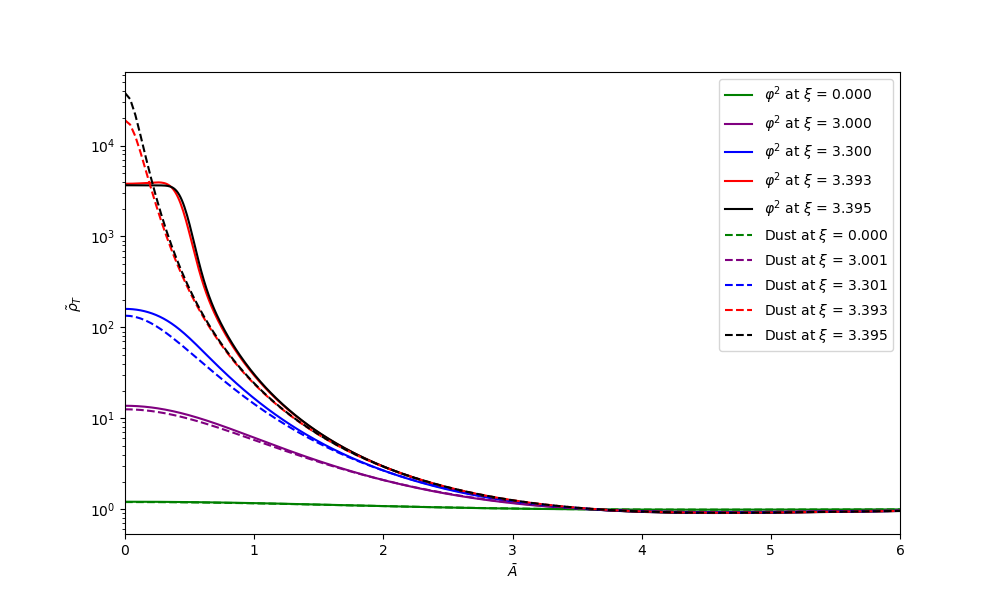}    \caption{\footnotesize{\textbf{Top-panel:} Evolution of the dimensionless central density for dust and scalar field perturbations. \textbf{Bottom panel:} Dimensionless density profile for dust and scalar field perturbations at different times.}}
    \label{fig:phi2}
\end{figure}

A key question we aim to address is whether if this effective pressure is observable in our simulations. To address this issue, we evolve the same type of initial conditions (with $\kappa = 0.2$) for both dust-like and scalar field perturbations. In Figure \ref{fig:phi2}, we illustrate the evolution of density in both configurations as a function of time. Notably, although the scalar field density initially appears to evolve similarly to dust perturbations (see the top panel of the figure), the bottom panel reveals that, once the quantum pressure kicks in, the central region of the density in the quadratic case becomes flattened. 

\begin{figure}
    \centering
    \includegraphics[width=8.5cm, height=12cm]{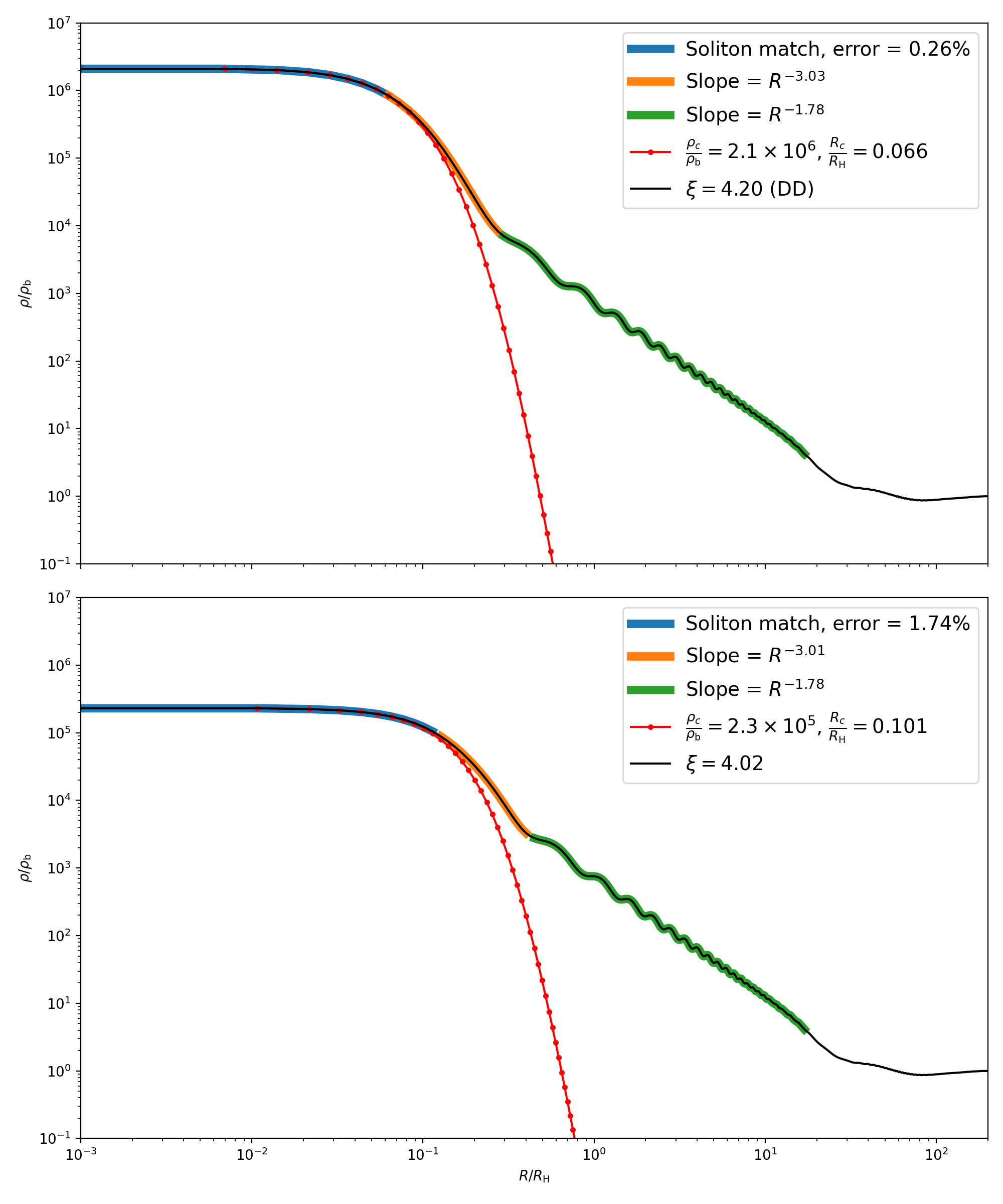}
    \caption{\footnotesize{The energy density as a function of \( R \) is shown for two different values of the logarithmic time, $\xi$. The central region of the numerical results are well fitted by the soliton solution (red dashed line). Beyond a certain radius, the profile deviates from the soliton solution and begins to decay with a slope of \( R^{-3} \) (blue curve). It then transitions into a profile with a slope of \( R^{-1.78} \) (black curve), finally reaching the cosmological background.
 }}
    \label{fig:sol_NFW}
\end{figure}

To gain a better understanding of the effects of this quantum pressure on the evolution of our perturbations, in Fig.~\ref{fig:sol_NFW} we plot the energy density as a function of the radius $R$, with both quantities expressed in physical units (rather than rescaled variables). The plot corresponds to two specific times before the formation of an apparent horizon, thus showing transition configurations. As shown in the figure, the central region of the density profile can be well approximated by the solution of a central soliton (depicted by the red dotted-dashed line){, with deviations not exceeding 2\% when compared to the analytical soliton profile.} For this soliton, we use the profile reported in \citep{2014NatPh..10..496S}, given by:
\begin{equation}
    \rho(R) = \frac{\rho_c}{(1+0.091(R/R_c)^2)^8},
\end{equation}
where $\rho_c$ is the central density and $R_c$ is the radius where the mass density drops by a factor of 2 from its value at the origin. Moreover, we observe that the decline of the central profile does not follow that of the soliton for large values of $R$. In fact, the profile decreases more gradually, transitioning from a falloff of $\propto R^{-3}$ (which closely resembles the NFW decay at large radii)\footnote{{The transition from the soliton core to the polynomial envelope was set at $R \simeq R_{c}$, in accordance with the behavior observed in Newtonian simulations.}} to a subsequent power-law decline  $\propto R^{-1.78}${with oscillations}, before asymptotically settling into the cosmological background. Let us emphasize that, to our knowledge, although the formation of solitons with NFW-like envelopes has been previously reported as an outcome of cosmological scalar field collapse (both, in spherical and 3D symmetries), this is the first time such a structure has been reported within the full general relativistic description.

By closely examining Fig.~\ref{fig:sol_NFW}, we observe that the profile of the central soliton undergoes a transformation over time: it becomes thinner while its central density increases. This behavior is consistent with numerical results in general relativity for \JCH{solitonic equilibrium configurations}, where it is well known that solitons with smaller radii tend to be denser (see, for example, \citep{Guzman:2009xre})\footnote{In fact, in the Newtonian regime, the relation 
$M_{\rm sol}R_{\rm sol} = $ constant holds.}.
Recall that this represents only an intermediate stage in the numerical evolution, as the perturbation ultimately collapses into a PBH. Nevertheless, this result provides insight into the nonlinear evolution of scalar field perturbations. Specifically, as the perturbation collapses, the quantum pressure arising from the wave-like nature of the scalar field allows the field to condense in the central region, forming a solitonic structure. As the fluctuation evolves, more matter accumulates in the central region, causing the soliton to grow in mass while shrinking in size. Eventually, once a critical amount of matter concentrates, the quantum pressure can no longer counterbalance self-gravity, leading to the perturbation’s gravitational collapse and the formation of a PBH. 

From the previous discussion, it is evident that a threshold value must exist, below which the matter accumulating in the central region of the distribution is insufficient to overcome the repulsive effects induced by the quantum pressure of the scalar field. In  this case, instead of forming a PBH, the system would stabilize into a virialised configuration, specifically a soliton with polynomial decay. Considering results from structure formation in the Newtonian regime, in Ref.~\citep{Padilla:2021zgm} it was determined threshold values at which either the soliton alone or the entire soliton + envelope structure collapses to form a PBH. The threshold values found in that study were:

\begin{equation}\label{eq:th} \delta^{\rm sol}_{\rm th} = 0.019, \quad \delta^{\rm sol+NFW}_{\rm th} = 0.238. \end{equation}

\begin{figure}
    \centering
    \includegraphics[width=9cm, height=5.5cm]{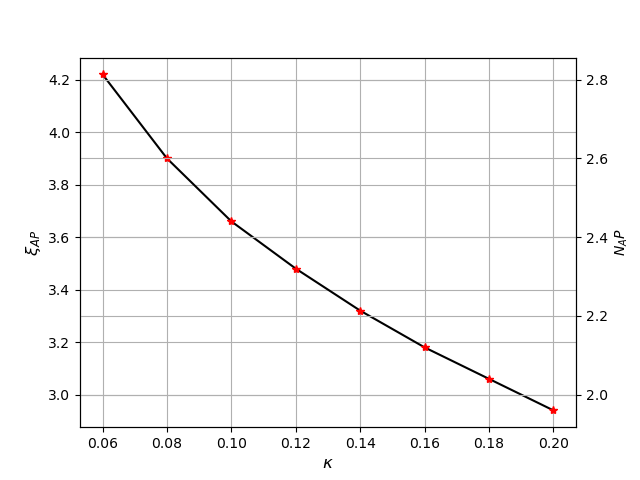}
    \caption{\footnotesize{Time interval from horizon entry until apparent horizon formation, as a function of different initial perturbation amplitudes $\kappa$. The left vertical axis displays logarithmic time $\xi_{AP}$ and the right shows the number of $e$-folds $N_{AP}$.  Points are the values we determined numerically.}}\label{fig:timeofcollapse}
\end{figure}

\begin{figure}
    \centering
    \includegraphics[width=9cm, height=5.5cm]{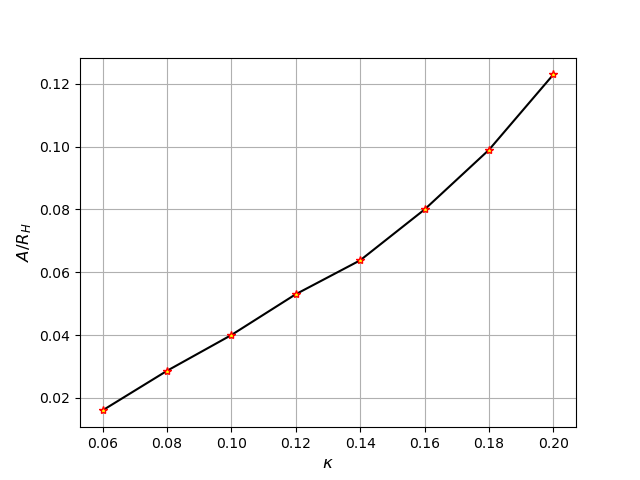}
    \caption{\footnotesize{The spatial coordinates at which the apparent horizon forms against different initial perturbation amplitude $\kappa$.}}\label{fig:kappasize}
\end{figure}

Building on the previous discussion and in the spirit of testing the accuracy of the threshold values reported above,\footnote{It is important to note that the gauge used to estimate the Newtonian results does not necessarily coincide with the one employed in this article, as we are working within a comoving frame tied to a radiation fluid.} we further analyze the evolution of different perturbations, considering various initial conditions for $\kappa$ within the range $\kappa \in \lbrace 0.06, 0.2 \rbrace$. As shown in Fig.~\ref{fig:timeofcollapse}, we observe that smaller initial amplitudes collapse at later times, with the time of collapse defined as the moment when the apparent horizon forms. Note that Fig.~\ref{fig:kappasize} demonstrates that smaller initial conditions result in smaller values at which the apparent horizon forms, \JCH{ relative to the Hubble radius}. This is consistent with the properties previously observed. 

{For initial amplitudes below $\kappa = 0.06$, we did not obtain simulations sufficiently reliable, according to our consistency tests, and within the limitations of the computational power available. One of the key challenges in this regime is that, as the initial amplitude is reduced, the scalar field undergoes an exponentially increasing number of oscillations before collapse. This significantly increases the runtime and imposes stringent demands on both spatial resolution and numerical stability. In particular, tracking the highly oscillatory evolution of the scalar field over extended timescales complicates maintaining control over the Hamiltonian constraint. Despite applying regridding techniques and increasing the resolution (from $N=800$ up to $N=10,000$), the magnitude of our normalized $\mathcal{H}$ eventually exceeds $\sim10^{-2}$ for very small amplitudes. This saturates the upper limit of acceptable numerical error. Therefore, in this work we restrict our analysis to the robust simulation regime $\kappa \in [0.06, 0.2]$, where we can ensure the numerical accuracy and physical reliability of our results.}
%Conversely, for initial amplitudes below 0.06, we have not obtained simulations with sufficient confidence in their physical validity \notesE{This will be answered in the Hamiltonian section}. As the initial amplitude decreases, the runtime increases exponentially due to the correspondingly exponentially increasing number of scalar field oscillations. Our current study focuses on the robust simulation regime between initial amplitudes of $\kappa=0.06$ and $\kappa=0.2$, where we have high confidence in the results. As of this work, we do not extend our analysis to the regime of very small-amplitude perturbations because the scalar field's evolution in this regime exhibits oscillations with exponentially growing amplitude over time. This makes it particularly challenging to accurately track such evolutions over the extended timescales and increasingly higher spatial resolutions required, given the highly oscillatory nature of the dynamics.

%Due to the fact that the evolution of the scalar field exhibits oscillations with exponentially growing amplitude over time, we do not explore the regime of small-amplitude perturbations. This is because accurately tracking their evolution over long timescales—especially given the scalar field’s oscillatory behavior—poses significant challenges.

\begin{figure}
    \centering
    \includegraphics[width=9cm, height=14cm]{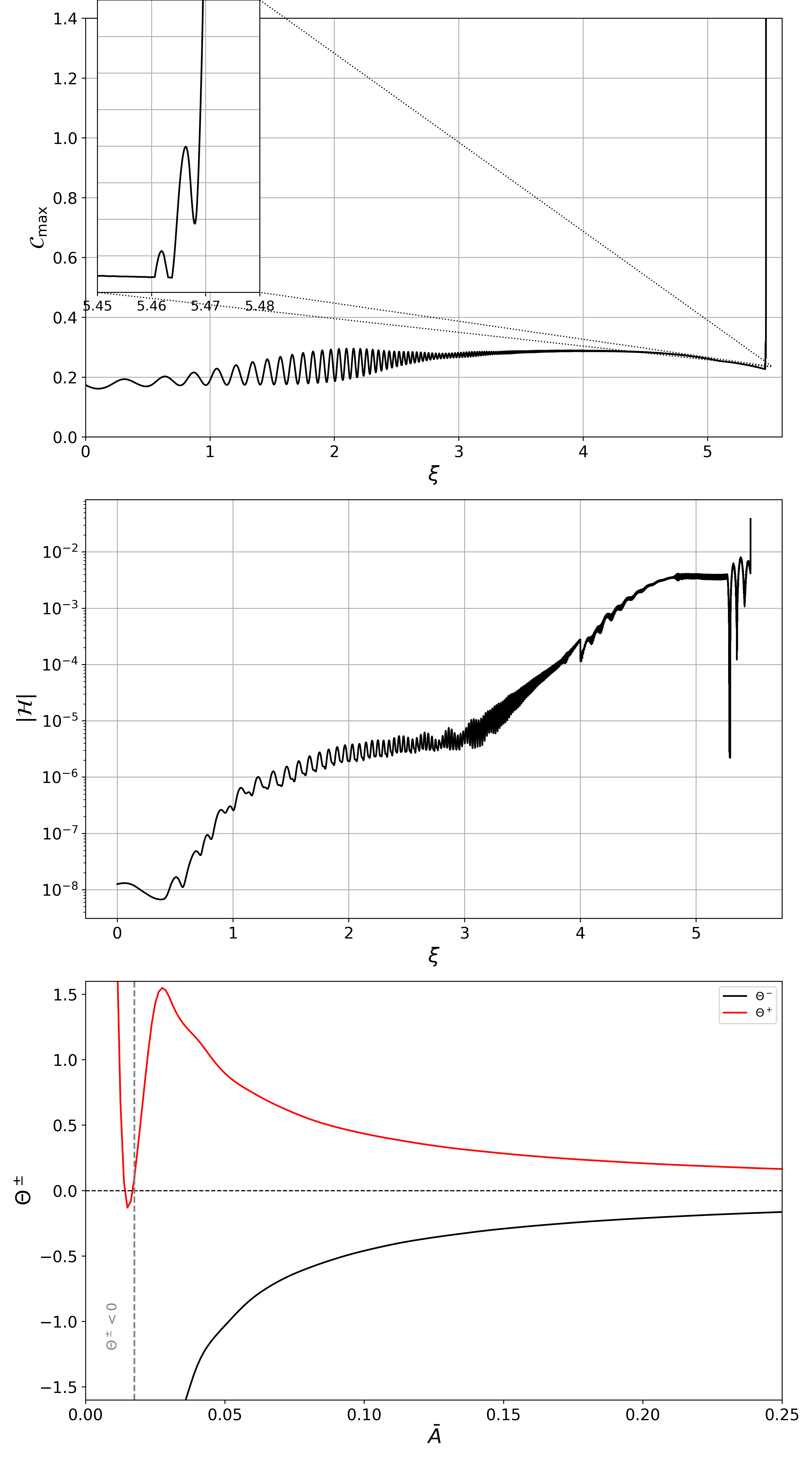}
    \caption{\footnotesize{\textbf{Top-panel:} Evolution of the maximum of the compaction function for the particular case of $\kappa = 0.06$. \textbf{Bottom panel}: Final time slicing, with a vertical line marking the location where the outgoing radial null geodesic vanishes, satisfying the condition $\Theta^{\pm}<0$ for a future trapped horizon. The condition that $\Theta^+=0$ and $\Theta^- < 0$ identifies the outermost trapped surface; the apparent horizon
 }}
    \label{fig:SlowReheatingAP}
\end{figure}

Across this range, we observe similar behavior: the perturbation initially collapses, and at a certain point, the central density begins to flatten before an apparent horizon forms. This recurring feature indicates that, within the framework of a collapsing quadratic scalar field, we can establish only an upper bound on the threshold necessary for PBH formation, which is associated with the minimum $\kappa$ value we evolved numerically. Specifically, our results indicate:  
\begin{equation}
    \mathcal{C}_{\rm th} \leq 0.2, \quad \delta_{\rm th} \leq 0.077.
\end{equation}  
The evolution of the maximum of the compaction function, {the mean value of the normalized expression for our hamiltonian consistency test}, and the appearance of an apparent horizon for this particular constraint is shown in Fig.~\ref{fig:SlowReheatingAP}. It is important to note that, although our simulations appear to predict the formation of a central soliton + NFW envelope structure, we find that these configurations still collapse into PBHs even for perturbation amplitudes smaller than those predicted by Newtonian estimates. This implies that the second reported threshold value in Eq.~\eqref{eq:th} is not accurate enough to reliably estimate PBH formation abundances in these scenarios, allowing PBHs to form at lower amplitudes than previously expected. 

Future work is needed to determine with more accuracy the threshold value for the collapse of inflaton stars.  

\section{Discussion and conclusions}\label{Sec:VII}

In this paper, we have developed and presented a fully relativistic numerical code based on the Misner-Sharp formalism for spherically symmetric cosmological scenarios containing both a scalar field and a perfect fluid. Our focus has been on cases where the scalar field dominates the dynamics, allowing us to explore the evolution of perturbations in scenarios such as slow-reheating after inflation.

We validated our numerical implementation by reproducing known results in the (perfect fluid) radiation-dominated scenario. Our results closely match the threshold for PBH formation in this regime, with a critical compaction function of approximately $\mathcal{C}_{\rm th} = 0.5$, consistent with previous works. The ability of our code to reproduce these results confirms the reliability of our approach.

We explored the case of a quartic potential for the scalar field, which is known to mimic the behaviour of a radiation-dominated universe in perturbative scenarios. As expected, we found that the PBH formation threshold is, up to the current resolution of our code, equal to that of the radiation-dominated case. This confirms the consistency of the analogy between a quartic scalar field potential and a radiation-dominated universe, even in a non-linear, strong gravity regime.

Interestingly, we find that the quadratic potential case deviates significantly from dust-like behaviour in the non-linear regime. Our simulations indicate that overdensities experience an effective quantum pressure that counteracts gravitational collapse. However, due to \JCH{limitations in computational power}, we were unable to evolve fluctuations with sufficiently low amplitudes, which in turn prevented us from determining a precise threshold for PBH formation in this case, leaving us only with an upper bound. Nevertheless, the behaviour of the perturbations we evolved suggests that PBH formation may differ significantly from the pure dust case \JCH{for which we have not managed to find the formation of an apparent horizon}, with a reasonable possibility that stable solitonic/virialised structures could emerge as the final outcome of gravitational evolution for some initial conditions.

{A key limitation preventing us from exploring smaller-amplitude perturbations is the difficulty in maintaining numerical control over the highly oscillatory dynamics of the scalar field for small inhomogeneities in this case. The exponential increase in the number of field oscillations, combined with the steepening of metric gradients, requires high spatial resolution and computational resources. Even after applying regridding techniques and using high-resolution grids, we find that the Hamiltonian constraint control expression eventually exceeds acceptable bounds (on the order of $10^{-2}$), marking the limit of numerical reliability. This constraint violation reflects not a breakdown of the physical dynamics, but rather the onset of numerical instability. As a consequence, we restrict our analysis to a regime where the simulations remain stable and accurate, enabling a robust assessment of PBH formation.}

In summary, our study provides new insights into PBH formation in scalar field-dominated scenarios using a fully relativistic numerical approach. While PBHs readily form in a quartic potential scenario, the quadratic case suggests the possible emergence of solitonic structures instead of PBHs. These findings highlight the importance of considering wave-like effects and quantum pressure in scalar field cosmologies.

\JCH{In previous works, the results of analytic approximations to the scalar field collapse, in terms of  associated threshold amplitudes for PBH formation, have been employed in constraining inflationary models with two characteristics: extended reheating phases, and features in the potential towards the end of inflation \cite{Hidalgo:2022yed,Padilla:2023lbv,Padilla:2024cbq}.  The implications for probes of the extended reheating scenario could be recast in light of the results of the present paper. On one hand, we are now confident that the constraints of the standard instantaneous reheating are readily applicable to the $\phi^4$, self-interacting case. On the other hand, the $\phi^2$ case is still not conclusive and deriving constraints may require information from the number of $e$-folds and energy scale of the reheating period.
Future work should explore these effects in greater detail, looking into more general scalar field models, as well as considering simulations with more than one matter component, phase transitions, and isocurvature modes sourcing the inhomogeneities. All these issues shall be explored elsewhere.}

{Furthermore, our results may have implications beyond the study of PBHs. Scalar fields are frequently considered as viable dark matter candidates, particularly in the context of ultra-light or self-interacting scalar field models. Although resolving the small-amplitude perturbations relevant for structure formation in these scenarios remains a major numerical challenge, our fully relativistic simulations offer valuable insights into the nonlinear dynamics of scalar fields under gravitational collapse. In particular, they may shed light on the possible formation of massive or even supermassive black holes in scalar field dark matter models, where solitonic cores and wave-like effects play a crucial role in the formation of these objects (see for example \citep{Padilla:2020sjy}).}

\section{Acknowledgments}
The authors would like to thank Katy Clough for many illuminating discussions. EM would also like to Charles Dalang for many useful discussions. EM is supported by an STFC studentship. LEP is supported by a Royal Society funded post-doctoral position. DJM was supported by a Royal Society University Research Fellowship for the majority of this work, and acknowledges current financial support from the STFC under grant ST/X000931/1. JCH acknowledges support from the UNAM-PAPIIT grant IG102123 “Laboratorio de Modelos y Datos (LAMOD) para proyectos de Investigación Científica: Censos Astrofísicos”, as well as from SECIHTI (formerly CONAHCYT) grant CBF2023-2024-162 and DGAPA-PAPIIT-UNAM grant IN110325 “Estudios en cosmología inflacionaria, agujeros negros primordiales y energía oscura.”
\appendix
\section{Obtaining the Cosmological Background}\label{app:background}

In this appendix, we derive the background equations of motion for both, the Misner-Sharp formalism described in Sec.~\ref{Sec:MS} and the cosmological decomposition given in Sec.~\ref{Sec:msc}.

\subsection{The Background Universe from the Misner-Sharp Formalism}

For the background universe, we impose the conditions $\rho_{\rm pf}(A,t) = \rho_{\rm pf,b}(t)$, $\varphi(A,t) = \varphi_{\rm b}(t)$, and $\Pi(A,t) = \Pi_{\rm b}(t)$, where suffix `$\rm b$' is used to refer to background quantities. From the above conditions, it immediately follows that $\chi = \chi_{\rm b} = 0$ and also
\begin{subequations}\label{eq:dapbA}
    \begin{equation}
        \rho_{\varphi,\rm{b}} = \frac{\Pi^2_{\rm b}}{2}+V(\varphi_{\rm b}),
    \end{equation}
    \begin{equation}
        P_{\varphi,\rm{b}} = \frac{\Pi^2_{\rm b}}{2}-V(\varphi_{\rm b}),
    \end{equation}
\end{subequations}
then $\rho_{\varphi,\rm{b}}$ and $P_{ \varphi,\rm{b}}$ are only time dependent. From Eq.~\eqref{eq:rhopf}, we obtain $\phi^{'}_{\rm b} = 0$, which under appropiate boundary conditions results in $\phi_{\rm b} = 0$. The remaining equations in Eqs.~\eqref{eq:MS} reduce to
\begin{subequations}\label{eqs:A2}
    \begin{equation}\label{eq:pibA}
        \dot\varphi_{\rm b} = \Pi_{\rm b},
    \end{equation}
    \begin{equation}
        \dot\Pi_{\rm b} +\left(\frac{2U_{\rm b}}{R_{\rm b}}+\frac{U^{'}_{\rm b}}{R^{'}_{\rm b}}\right)\Pi_{\rm b} +V(\varphi_b)_{,\varphi_b} = 0.
    \end{equation}
    \begin{equation}
        \dot R_{\rm b} = U_{\rm b},
    \end{equation}
    \begin{equation}
        \dot m_{\rm pf,b} + 4\pi R_{\rm b}^2 P_{\rm pf,b}U_{\rm b} = 0,
    \end{equation}
    \begin{equation}
        \dot U_{\rm b} = -\left(\frac{M_{\rm T,b}}{R_{\rm b}^2}+4\pi R_{\rm b}P_{\rm T,b}\right)
    \end{equation}
    \begin{equation}\label{eq:rhopfbA}
        \rho_{\rm pf,b} = \frac{m^{'}_{\rm pf,b}}{4\pi R_{\rm b}^2 R^{'}_{\rm b}}
    \end{equation}
    \begin{equation}\label{eq:rhophibA}
        \rho_{\varphi \rm,b} = \frac{m^{'}_{\varphi \rm,b}}{4\pi R_{\rm b}^2 R^{'}_{\rm b}}
    \end{equation}
    \begin{equation}\label{eq:Gam2bA}
        \Gamma_{\rm b}^2 = 1+U_{\rm b}^2 - \frac{2m_{\rm T,b}}{R_{\rm b}}.
    \end{equation}
\end{subequations}

To simplify the above system, we can immediately solve Eqs.~\eqref{eq:rhopfbA} and \eqref{eq:rhophibA} to obtain $M_{\rm pf,b} = 4\pi \rho_{\rm pf,b} R_{\rm b}^3/3$ and $M_{\varphi \rm,b} = 4\pi \rho_{\varphi\rm,b} R_{\rm b}^3/3$. The fact that in the rest of the equations there is not spatial derivatives imply that the evolution of each function will be independent of $A$. In particular, we can express $R_{\rm b}$ as $R_{\rm b} = a(t)A$, where $a(t)$ is the usual scale factor. We parameterized the coordinate system such that for $t = t_0$, $a(t_0) = 1$. Using the above expressions, defining the Hubble parameter $H$ as $H \equiv \dot a/a$, and after some algebraic manipulations, we reduced Eqs.~\eqref{eq:pibA} to \eqref{eq:rhophibA} to the following relevant differential equations:
\begin{subequations}
\begin{equation}
    \ddot\varphi_{\rm b} +3H\dot\varphi_{\rm b}+V(\varphi_{\rm b})_{,\varphi_{\rm b}} = 0,
\end{equation}
\begin{equation}\label{eq:pfbA}
    \dot\rho_{\rm pf,b}+3H(\rho_{\rm pf,b}+P_{\rm pf,b}) = 0,
\end{equation}
\begin{equation}
    \frac{\ddot a}{a} = -\frac{4\pi}{3}(\rho_{\rm T,b}+3P_{\rm T,b}),
\end{equation}
i.e. we obtain the usual KG, continuity, and acceleration equation for a cosmological background. The KG equation can be further manipulated to obtain
\begin{equation}\label{eq:cphibA}    \dot\rho_{\varphi\rm,b}+3H(\rho_{\varphi\rm,b}+P_{\varphi\rm,b}) = 0,
\end{equation}
i.e. we arrive at a continuity-like equation valid for the scalar field. However, it is important to clarify that the scalar field is not a perfect fluid, as it is generally not possible to write an equation of state in the form of $P_{\varphi\rm ,b} = w_\varphi\rho_{\varphi\rm ,b}$. Instead, the density and pressure evolve according to Eqs.~\eqref{eq:dapbA}. Combining the above three equations, it is straightforward to arrive at the Friedmann equation:
\begin{equation}\label{eq:fbA}
    H^2 = \frac{8\pi}{3}\rho_{\rm T,b}-\frac{k}{a^2},
\end{equation}
where $k$ is an integration constant that is related with the curvature of the background universe. We can multiply the above equation by $R_{\rm b}^2$, substitute for $U_{\rm b}$, and use Eq.~\eqref{eq:Gam2bA} to obtain
\begin{equation}
    \Gamma_{\rm b}^2 = 1-kA^2.
\end{equation}
Then, the metric in the Misner-Sharp formalism reduces to
\begin{equation}
    ds^2 = -dt^2+a(t)^2\left(\frac{dA^2}{1-kA^2}+A^2d\Omega^2\right),
\end{equation}
\end{subequations}
where we used $e^\lambda = R^{'2}/\Gamma^2$. The above expression is basically the FLRW metric, used to describe the background universe.

Before concluding this section, let us highlight the important result of having Eq.~\eqref{eq:cphibA}. Combining Eq.~\eqref{eq:pfbA} and \eqref{eq:cphibA}, we obtain
\begin{equation}\label{eq:rhotA}    \dot\rho_{\rm T,b}+3H(\rho_{\rm T,b}+P_{\rm T,b}) = 0.
\end{equation}
This means that if we can write the total pressure of the system as $P_{\rm T,b} = w\rho_{\rm T,b}$, where $w$ is an effective equation of state of the complete system, we can immediately solve the above equation and, in turn, the Friedmann equation, Eq.~ \eqref{eq:fbA}. Although in general it is not possible to obtain such an effective equation of state, for the scenarios studied in this work it is indeed possible to have an effective $w$. For example, in the case where the perfect fluid dominates, we have $w = w_{\rm pf}$, while in the case in which a fast oscillating scalar field in a polynomial potential dominates, Eq.~\eqref{eq:quadratic}, we have an averaged equation of state given by $w = (n-1)/(n+1)$.

\subsection{The Background Universe from the Cosmological Misner-Sharp Decomposition}

Our intention now is to obtain the background dynamics from Eqs.~\eqref{eq:final}. As a first step, it is convenient to remember the different variables introduced in \eqref{eqs:cosmoMD}.
Comparing with our background solutions in the above section, it is evident that the background universe implies: 
\begin{subequations}
\begin{equation}
    \tilde R_{\rm b},\
    \tilde U_{\rm b}, \
\tilde m_{\rm T,b} = 1,\ \ \ \text{and} \ \ \ \tilde\chi_{\rm b} = 0.
\end{equation}
\end{subequations}
From the previous section, we had also found that $\phi_{\rm b} = 0$. Considering these simplifications, we can reduce Eqs.~\eqref{eq:final} to the following relevant differential equations:
\begin{subequations}
    \begin{equation}\label{eq:vphibA}
        \partial_\xi\tilde \varphi_{\rm b} = \alpha\left[e^\xi\tilde \Pi_{\rm b}+\frac{3(\rho_{\rm T,b}+P_{\rm T,b})}{2\rho_{\rm T,b}}\tilde\varphi_{\rm b}\right],
    \end{equation}
    \begin{equation}\label{eq:pibA}
        \partial_\xi\tilde\Pi_{\rm b} = \alpha\left[\frac{3}{2}\left(\frac{P_{\rm T,b}}{\rho_{\rm T,b}}-1\right)\tilde\Pi_{\rm b}-e^{\xi}\tilde{V}(\tilde{\varphi_{\rm b}})_{,\tilde{\varphi_{\rm b}}}\right],
    \end{equation}
    \begin{equation}\label{eq:mpfA}
        \partial_\xi \tilde m_{\rm pf,b} = 3\alpha\left(\frac{P_{\rm T,b}}{\rho_{\rm T,b}}\tilde m_{\rm pf,b}-\tilde P_{\rm pf,b}\right).
    \end{equation}
\end{subequations}
At this point, we can continue providing general results from the previous system. However, we will limit ourselves to discussing only the two scenarios of interest in this work: perfect fluid and scalar field domination.

\subsubsection{Perfect fluid-dominated scenario}

For the perfect fluid dominated scenario, we set $\tilde\varphi_{\rm b} = \tilde\Pi_{\rm b} = 0$. We also consider $\tilde m_{\rm pf,b} = \tilde m_{\rm T,b} = 1$, which from Eq.~\eqref{eq:mpfA} implies that $\tilde P_{\rm pf,b} = w$, where $w$ is the equation of state of the dominant fluid, i.e. $w = w_{\rm pf}$. Then, it is straightforward to obtain that $\tilde \rho_{\rm pf,b} = 1$.

\subsubsection{Scalar field dominated scenario}

For this scenario, we consider that all the variables related with the perfect fluid are equal to zero. Then, the only relevant equations to be solved are Eqs.~\eqref{eq:vphibA} and \eqref{eq:pibA} which, in general, must be solved numerically once a particular form for the scalar field potential $V(\tilde\varphi_b)$ is adopted. In order to have an analytical solution to compare with our numerical results, in this section we shall reduce ourselves to the simple quadratic potential ($n = 1$ and $\lambda_1 = \mu^2$) in Eq.~\eqref{eq:quadratic}. To simplify the system further, we also assume $\alpha = 2/3$ and $P_{\rm T,b}\simeq 0$, i.e. we consider the mean background behaves as a matter-dominated universe. Such a behaviour is consistent with the quadratic potential scenario we are considering in this section. Combining Eqs.~\eqref{eq:vphibA} and \eqref{eq:pibA} and after doing some algebraic manipulations, we obtain
\begin{equation}
    \partial^2_\xi \tilde\varphi_{\rm b} - \partial_\xi\tilde\varphi_{\rm b}+\left(\frac{2}{3}e^\xi\tilde\mu\right)^2\tilde\varphi_{\rm b} = 0,
\end{equation}
where $\tilde \mu = R_H \mu$. The result of the above equations is:
\begin{equation}
    \tilde \varphi_{\rm b} = \tilde\varphi_{\rm 0,b}\cos\left(\frac{2}{3}\tilde\mu e^\xi+\delta_0\right),
\end{equation}
i.e. we obtain that the scalar field background in tilde variables oscillates in time with an exponential growing frequency. It is important to emphasize that this oscillatory behavior was obtained because we assumed that the cosmological background effectively behaved like a pressureless fluid, so the expected oscillatory dynamics of the scalar field were inherited by the tilde variables.

\bibliographystyle{ieeetr}
\bibliography{biblio}

\end{document}